\documentclass{article}
\usepackage[defaultsups]{newtxtext}
\usepackage[varg]{newtxmath}
\usepackage[superscript,nomove]{cite}
\usepackage{graphicx}

\usepackage{color}
\usepackage{soul}

\title{PMSM transient response optimization by end-to-end optimal control}
\author{Yuta Kawachi\thanks{DENSO IT Laboratory} \and Mitsuru Ambai\thanks{DENSO IT Laboratory} \and Yuichi Yoshida\thanks{DENSO IT Laboratory} \and Gaku Takano\thanks{DENSO IT Laboratory}}

\date{February 2024}

\begin{document}

\maketitle

\begin{abstract}
Speed responses of motors, especially Permanent Magnet Synchronous Motors (PMSMs), are increasing in importance for recent applications, such as electric vehicles or quadrotors. These applications require quick acceleration performance. However, commercial controllers are based mainly on Proportional-Integral (PI) controllers, which are suitable for eliminating steady-state errors but unsuitable for transient response optimization. In this paper, we replaced whole conventional controllers with an end-to-end Recurrent Neural Network (RNN) that has a regularized transition matrix. Our end-to-end controller directly minimizes the transient response time on the basis of optimal control theory. Computer-simulated results show that speed response indices improved using the RNN rather than a PI controller, while both were under comparable power losses. The current vector trajectories of the RNN showed that the RNN could automatically determine arbitrary trajectories in the flux-weakening region in accordance with an arbitrarily designed loss function. In contrast, the traditional flux-weakening methods using PI controllers have pre-determined current vector trajectories.
\end{abstract}

\section{Introduction}
\label{sec:int}
Motors are core components of electrification that will revolutionize the fossil fuel society. Applications of motors are various: from relatively classical applications such as factories or trains to cutting-edge, response-intensive applications such as autonomous cars or quadrotors. Autonomous cars run on unpredictable random roads. Quadrotors are required to track random orbits in the air quickly. In recent response-intensive applications, PMSMs (also called Direct Current Brush-Less Motors) are widely adopted. One of the dominate merits of using PMSMs over other classical brushed motors is that they are mechanically robust because there is no brush wear. However, this brushless property fundamentally requires alternating currents or switching mechanisms. 

The nature of alternating current makes PMSM responses complex. For torque-intensive applications, especially for EVs, the Interior Permanent Magnet Synchronous Motor (IPMSM) is more suitable than the Surface Permanent Magnet Synchronous Motor (SPMSM). However, as the IPMSM is the general case of the SPMSM, it is known to have more complex behavior than the SPMSM, which reflects the magnetic saliency of IPMSMs. The behavior is explained by the difference in the number of solutions of underlying non-linear ordinally differential equations (ODEs)\cite{li2002bifurcations, souhail20162d, liu2021dynamic, takougang2021spiking, qu2021bursting}. These properties suggest that PMSMs require sophisticated algorithms regarding the non-stationary, non-linear properties of the underlying ODEs.

A typical de-facto standard of commercial PMSM controllers is the PI controller based on Field-Oriented Control (FOC). {This is also called vector control}. This method is characterized by using on-magnet coordinates, which means direct and quadrature coordinates\cite{Park1928dq} of two-dimensional current vector ${\mathbf i}_{dq}$ space. PI controllers are mathematically justified by classical control theory. This theory assumes that a system is linear. If the system is non-linear, a local property is described by known linear techniques (e.g., linearization around the equilibria of a system), but modeling a global property requires non-linear techniques. If we model PMSM controllers using PI controllers in a broader operation range (e.g., higher, random; torque or speed), which usually causes non-linear effects, impractical numbers of PI controllers and tunings will be required.

In this paper, to tackle the non-stationary, non-linear nature of PMSMs, we introduce Recurrent Neural Network (RNN) controllers based on Machine Learning (ML). The ML approach is based on an inductive way of thinking. It is fundamentally different from the classical deductive idea, which assumes access to the analytical properties of the solutions of a system. Instead of (human) approximation and simplification of a system, the ML approach is a data-driven approach that automatically makes empirical models from some form of data. In comparison, the ML approach, which defines loss functions to be minimized, is similar to the idea of optimal control theory \cite{BOLTYANSKI1960464}. Optimal control theory also defines a functional over a trajectory, which should also be minimized. Both the differences and the similarities are suitable for automatic optimization without impractical human effort required for hand tuning. The initial approach to this problem can be done in two steps: solving the original non-linear optimal control problem straightforwardly to get the optimal control input function and learning a controller that maps states into control inputs. However, the former is known to be almost impossible due to analytical difficulty \cite{Yu2022Optimal}. Instead of this two-step idea, we fuse these approaches into one using the end-to-end idea of neural ODEs\cite{chen2018neural}. These approaches unroll (construct computational graphs corresponding to the discretization method of the ODEs, such as Euler or Runge-Kutta methods) the whole state feedback system, including the RNN controller and IPMSM plant. As RNNs have non-linear activation functions, such as a Rectified-Linear Unit (ReLU\cite{pmlr-v15-glorot11a}), the controller is also capable of treating non-linear phenomena of PMSMs. Training, the counterpart of the gain-tuning of PI controllers, is fully automatically done by stochastic gradient descent (SGD) methods based on the optimal control-based discretized functional loss representation. The usage of SGD enables the minimization of some energy indices (e.g., copper loss) and some transient indices (e.g., speed settling time). The latter is known to be problematic in classical control theory because the non-linear transient property is related to the time-domain solutions of the general non-linear system. However, the solutions of equilibria, which correspond to the stationary state, are relatively easy to treat.

We compared proposed RNN controller and conventional PI-FOCs as IPMSM speed controllers. We obtained a computer-simulated result that suggests that RNN well extrapolate to faster speed references in a shorter amount of acceleration time than PI-FOCs under comparable copper loss.

\section{Related work}
\label{sec:rel}
In academic or research fields, many controllers have been proposed in addition to PI-FOCs.

\subsection{Model predictive control (MPC)}
Model predictive control is based on optimal control theory and assumes known plant ODEs. There is a simulated result of MPC-based motor control\cite{9681029}. This is similar to our approach, but the most significant difference is the computational load coming from its model-predictive nature. MPC-based methods predict a plant's future response from now to some fixed time length (the prediction horizon) at each time step. This algorithm has serial iteration, which is difficult to parallelize for general non-linear systems. This property requires CPU specifications like PCs. Therefore, applications with long response times and computationally-rich controllers are suitable for MPCs. There are some examples of vehicle control applications for quadrotors \cite{bicego2020nonlinear, sun2022comparative}; however, their controllers are PCs or have GPU-like rich architectures. For our applications, such as the motors inside EVs or quadrotors, this is typically not the case. We concluded that these approaches are unsuitable for controlling motors that have a shorter response time than the vehicle and typically have cheaper hardware with stricter regulations regarding heat.

\subsection{Reinforcement learning (RL)}
As its computational load is not that high, reinforcement learning has been deployed for motor control itself\cite{khiabani2020optimal, Zhao2022Reinforcement, 9236258, 9334732, 9531457}. Recently, deep RL-based methods that correct the signals in FOC \cite{Nicola2021Improvement}, replacing the speed PI controller\cite{Wang2023Active} or current PI controllers\cite{Bhattacharjee2020Policy}, have been tested. Although these can be viewed as nature-inspired algorithms, we can also seek analogies between RL (in continuous space) and optimal control theory\cite{levine2018reinforcement}. For example, policy gradients \cite{sutton1999policy} have probabilistic counterparts of the state feedback system: actions, control inputs, state transition probabilities, plant models, policies, and controllers. Thus, RLs and our approach are similar, but the main difference is whether one of the plants or losses (corresponding to a reward) is known or not\cite{sasaki2019reinforcement, recht2019tour, ODonoghue2020Making}. Recently, standard RL algorithms have been regarded as expectation maximization\cite{fellows2019virel}. Thus, RL methods can be said to be a type of latent probabilistic optimal control. This means that the plant and/or the functional loss are assumed to be unknown under the optimal control problem for a probabilistic state feedback system. This property is suitable for tasks involving massive systems or human/animal decisions (e.g., human-computer interaction with languages \cite{brown2020language}), which are hard to describe directly. However, instead of modeling flexibility, RL methods are known to be challenging to train as complicated techniques are required \cite{Engstrom2020Implementation}. In our paper, all functions are set to be known: we arrange a known (identified) plant ODE and loss functions with known calculation methods (e.g., copper loss). This arrangement means we do not need to apply RL to this task; we need not pretend to be ignorant of these functions. These settings enable straightforward gradient descent-based training, and we can expect the training to be much easier.

\subsection{Neural networks (NNs)}
PMSMs are known to have couplings between the currents or current and speed, which are the state variables of the governing equation. These couplings between variables are one of the origins of the PMSM's non-linearity. NNs are deployed to eliminate these couplings because analytical decoupling methods are known to be insufficient. There are combinations of classical controllers and NN-based decoupling\cite{li2019neural, Li2018Decoupling, 9641002, 8595688}. Similar techniques are also used to model time-varying PI controllers\cite{Jie2020TimeVarying}. {NN-based controllers for SPMSM or $i_d=0$ control strategy has also been tested}\cite{9615146, 9957127, 9984109, 10023763, 9229198}.
However:
\begin{itemize}
 \item Time-independent NNs, such as Multi-Layer Perceptrons (MLPs) or Radial Basis Function (RBF) NNs \cite{9275116, 10023763, 9969086, 9615148}, have a limited dynamics representation capability. The dynamics are typically represented only through the traditional integration of the error signal.
 \item The non-end-to-end architectures, which means a pipeline-like system of NNs and PI controllers, 
 have a human design of the d-axis current reference ${\hat {i_d}}$ based on steady-state assumptions. 
 This design will lead to suboptimality of the transient speed response performance, especially for IPMSM, which has magnetic saliency ($L_d \neq L_q$), additional $i_d$ term in the torque equation that causes multiplicative nonlinearity between d- and q-axis current, and requires $i_d \neq 0$ control strategies.
 \item The NN controllers, which replace only some parts of the traditional controllers, face the problem of selecting training criteria.  If the NN subsystem is trained by the PI controller output, the performance will be bounded by the PI controller performance.
\end{itemize}
In this paper:
\begin{itemize}
 \item The time-dependent RNN has a nearly unlimited non-linearity capability corresponding to the number of hidden layers. 
 \item The end-to-end RNN architecture, which eliminates human ${\hat {i_d}}$ design, has no limitation on trajectories in the current vector space, which will improve the speed response time.
 \item Without PI controllers, the intuitive optimal control-based loss function can be used to train the RNN directly to minimize the loss. The loss design also simultaneously optimizes multiple criteria: the speed response and copper loss minimization.
\end{itemize} 

\section{Conventional method}
\label{sec:conv}
\subsection{PMSM plant}
We first explain the standard well-known magnetically-linear dq-axes non-linear PMSM ODE\cite{Yu2020Modeling} as 
\begin{align}
\label{eq:pmsm}
&\frac{{\rm d}}{{\rm d} t} 
\left[
\begin{matrix}
i_d \\
i_q \\
\omega_e \\
\end{matrix}
\right] \notag \\
&=
\left[
\begin{matrix}
-\frac{R}{L_d}i_d + \frac{L_q}{L_d}\omega_e i_q + \frac{1}{L_d}v_d \\
-\frac{L_d}{L_q}\omega_e i_d -\frac{R}{L_q}i_q + \frac{1}{L_q}v_q - \frac{\rm \Phi}{L_q}\omega_e\\
-\frac{D}{J}\omega_e + \frac{P^2}{J}({\rm \Phi} + (L_d - L_q)i_d)i_q - \frac{P}{J}T_L\\
\end{matrix}
\right]
\end{align}
where the variables, including states, control inputs, external forces, and constant parameters, are defined in Table \ref{table:vars}.
\begin{table}[tb]
\caption{Variable definition}
\label{table:vars}
\begin{center}
\begin{tabular}{c|c|c}
\noalign{\hrule height 0.4mm}
name & symbol & unit\\
\hline
d-axis current & $i_d$ & [A] \\
q-axis current & $i_q$ & [A]\\
electrical angular velocity & $\omega_e$ & [rad/s]\\
\hline
d-axis voltage & $v_d$ & [V] \\
q-axis voltage & $v_q$ & [V] \\
\hline
load torque & $T_L$ & [Nm]\\
\hline
winding resistance & $R$ & [$\rm \Omega$] \\
d-axis inductance & $L_d$ & [H] \\
q-axis inductance & $L_q$ & [H] \\
permanent magnet flux & ${\rm \Phi}$ & [Wb] \\
number of pole pairs & $P$ & [] \\
moment of inertia & $J$ & [$\rm kgm^2$] \\
viscous friction & $D$ & [$\rm Nm s/rad$] \\
\noalign{\hrule height 0.4mm}
\end{tabular}
\end{center}
\end{table}
Note that we may describe a time function $f(t)$ as $f$ without its time variable $t$. When $L_d = L_q$, this models SPMSM, and when $L_d \neq L_q$, this models IPMSM. This equation is non-linear because of the multiplicative non-linearity of the speed (electrical angular velocity) and the current, or that between the dq currents. Even if there is no magnetic non-linearity, such as hysteresis in the B-H plane, this model is known to cause chaotic behavior depending on the parameter values\cite{li2002bifurcations, takougang2021spiking}.

\subsection{PI-FOCs}
\begin{figure}[t]
\begin{center}
 \includegraphics[width=\linewidth]{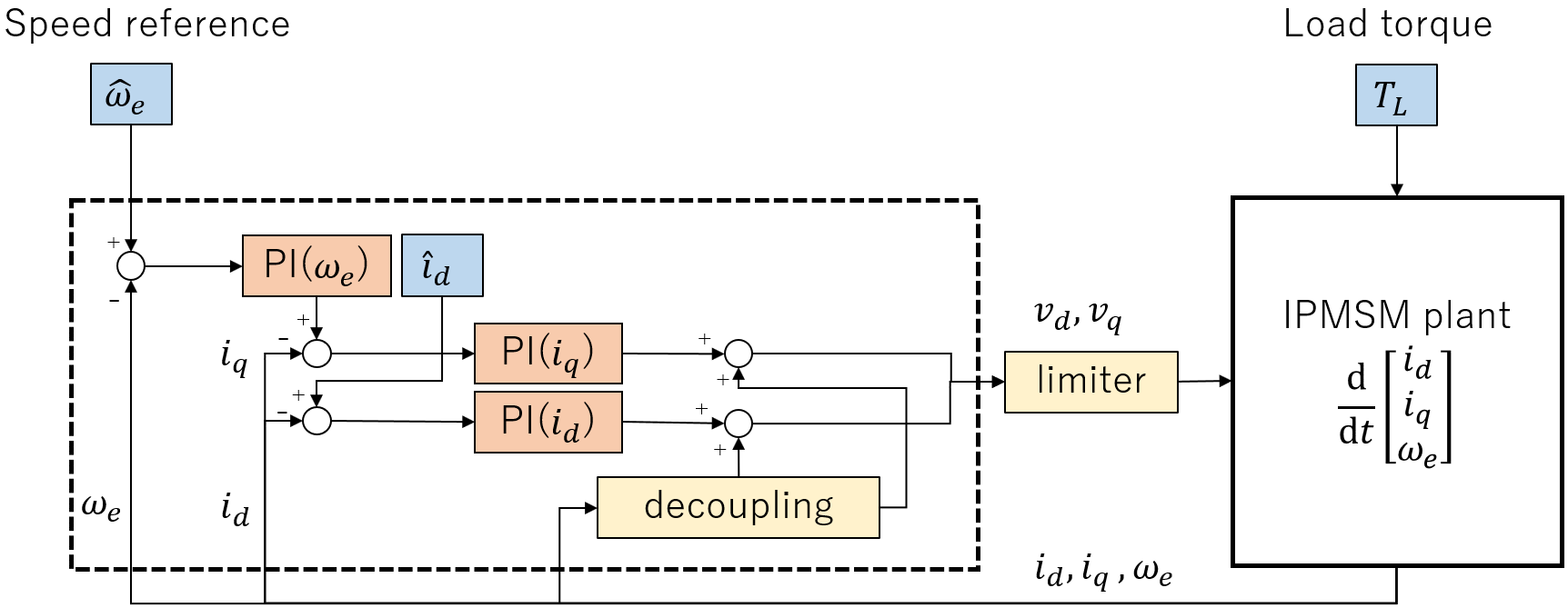}
\end{center}
\caption{Baseline PI-FOC system. Current reference limiters and anti-windup components are added if the baseline performance, depending on the simulation condition, is not enough. Note that the proposed RNN controller replaces all of the components inside the dotted line as a black-box.}
\label{fig:block_diagram}
\end{figure}

PI-based FOCs (PI-FOCs) are state feedback systems that are commonly used in commercial PMSM controllers. As they do not have computationally-intensive, especially iterative, algorithms, they are suitable for microcontrollers embedded in nearby PMSMs. In this paper, we consider the typical cascaded speed and current controller with decoupling and a voltage limiter feature. We denote reference speed and reference dq-axis currents as ${\hat \omega_e}$, ${\hat i_d}$, and ${\hat i_q}$. Also, we denote each PI gain as ${ k_{p\omega}}$, ${k_{i\omega}}$, ${k_{pd}}$, ${k_{id}}$, ${k_{pq}}$, and ${k_{iq}}$. The state feedback system can be obtained by substituting 
\begin{eqnarray}
\label{eq:FOC}
\frac{{\rm d}}{{\rm d} t} 
\left[
\begin{matrix}
s_{id} \\
s_{iq} \\
s_{\omega} \\
\end{matrix}
\right]&=&
\left[
\begin{matrix}
{\hat i_d} - {i_d} \\
{\hat i_q} - {i_q} \\
{\hat \omega_e} - {\omega_e} \\
\end{matrix}
\right]
\end{eqnarray}
\begin{eqnarray}
{\hat i_q} &=& {k_{p\omega}}\cdot({\hat \omega_e} - {\omega_e}) + {k_{i\omega}}s_{\omega} \nonumber \\
{v_d} &=& {k_{pd}}\cdot({\hat i_d} - {i_d}) + {k_{id}}s_{id} - L_q i_q \omega_e \nonumber \\
{v_q} &=& {k_{pq}}\cdot({\hat i_q} - {i_q}) + {k_{iq}}s_{iq} + {\rm \Phi} \omega_e + L_d i_d \omega_e \nonumber \\
\end{eqnarray}
into the original plant Eq. (\ref{eq:pmsm}). The equivalent block diagram of the resulting control system is shown in Fig. \ref{fig:block_diagram}.

Even when combined with the controller, the state feedback system still is known to cause chaotic behavior\cite{souhail20162d}. This fact implies that PMSMs are complex and that more sophisticated controllers are essentially needed.

\subsection{Flux weakening (FW) methods}
\label{subsec:fw}
As the d-axis is the N pole direction of the magnet, decreasing the d-axis current to negative is similar to weakening the magnetic flux, which can be done to increase the speed limit, but determining the reference d-axis current ${\hat i_d}$ is not straightforward. This is the main focus of FW strategies. On the basis of Eq. (\ref{eq:pmsm}), steady state dq current relationships called the Maximum Torque Per Ampere (MTPA)\cite{Jahns1986MTPA}, Maximum Torque Per Flux (MTPF)\cite{Morimoto1990MTPF}, and Maximum Torque Per Voltage (MTPV) are derived\cite{Lu2010fw, Christoph2017Analytical, eldeeb2018unified, Ekanayake2018MTPV} from a copper loss minimization criterion or inverter voltage ratings. In addition, the current limit forms a circle in the dq current space. We call this Maximum Current (MC) in this paper. These relationships in the $i_d < 0$ subspace are used in the FW methods in a broad sense. In particular, the normal operation region surrounded by the MTPA, MTPV, and MC is called the flux-weakening region. In a narrow sense, an FW method must set the operation point on a voltage limit ellipse, which is determined by speed. We call this the Maximum Voltage (MV). The combination of MV and MC equals the Maximum Power (MP). However, an MP control strategy is purely feed-forward and does not have a speed compensation mechanism. This means a subtle error in the system identification directly affects the speed steady state error. This property is generally not good for practical applications. For this reason, we use a PI controller for speed, and the remaining $i_d$ is determined from one of the ${\mathbf i}_{dq}$ relationships described already. This is the baseline PI-FOC. If we select a dq current relationship, we can determine ${\hat i_d}$ from ${\hat i_q}$. Commercial FW controllers are usually a combination of these relationships conditioned on the current state; however, the possible current trajectory patterns are very limited on these curves, and a current reference occasionally discontinue when mode switching happens. This may cause unwanted responses or noise. 

\subsection{Selection of speed reference function}
\label{sec:ramp}
We mainly aim at optimizing transient speed responses. As the form of the speed reference function ${\hat \omega_e}$ limits the speed responses, the selection of the reference is worth discussing. At startup, instead of step functions, saturated ramp functions are empirically used in the speed reference of PMSMs. For some vehicle applications with payloads or passengers, the maximum acceleration exists, and the step speed function is unsuitable. The actual reason for this selection is unclear, but a possible reason comes from the non-linearity of PMSMs. As there are multiple equilibria in PMSMs, there are the cases that a desirable equilibrium is located beyond the separatrix from the current state when using a step reference function (which means the control inputs are a fixed parameter). If we use infinitely-flat ramp functions as the reference, which can be said to be the same as moving the equilibrium very slowly, we can expect the speed to reach the target speed corresponding to the target equilibrium. Thus, generally, while a large slope similar to the step function is preferable from the viewpoint of response time, a small slope is better in terms of control.
Another reason can be found from a (brushed) DC motor model, which can be viewed as a more simplified SPMSM. As the DC motor 
\begin{equation}
\label{eq:dcm}
\frac{{\rm d}}{{\rm d} t} 
\left[
\begin{matrix}
i \\
\omega \\
\end{matrix}
\right]
=
\left[
\begin{matrix}
 - \frac{Ri}{L} +\frac{v}{L}- \frac{{\rm \Phi}\omega}{L} \\
\frac{{\rm \Phi} i}{J} - \frac{T_L}{J}\\
\end{matrix}
\right]
\end{equation}
(subscript of variables omitted because there are no differences with DC motor) is linear, we can assume that the electrical response is faster than the mechanical response. Thus, we assume electrical stationarity $\frac{{\rm d}i}{{\rm d} t}=0$; then, the speed can be solved as an exponential form having the time constant $RJ/{\rm \Phi}^2$. From this discussion, the reference should at least be a saturated exponential function, and the usage of the saturated ramp function is a low-speed approximation of the saturated exponential function. However, we do not think the saturated exponential is enough because:
 \begin{itemize}
 \item The IPMSM has magnetic saliency ($L_d \neq L_q$) and coupling, which changes the analytical properties largely.
 \item The moment of inertia is difficult to determine in in-the-wild situations.
 \end{itemize}
Thus, the speed reference should be flexibly (re-) determined by using methods based on some kind of data-driven technique.
 
\section{Proposed method}
\label{sec:intention}
\subsection{Controller design policy}
As we mentioned in Sect. \ref{subsec:fw}, FW methods, based on a steady-state assumption, have some drawbacks even for traditional steady-state applications due to the limited current trajectory selections. Moreover, recent applications are operated under highly non-stationary conditions. The core idea of our controller design is unbinding the current trajectories from pre-specified current references (e.g., MTPA) in the current vector control. As we aim at optimizing the controller to minimize the transient speed responses and the energy criterion at the same time, the corresponding optimal current trajectories are generally unknown. This fact requires controller flexibility to determine the current vector arbitrarily. We use an RNN controller, optimal-control-based loss function, and SGD-based training method to automatically approximate the optimal current trajectories.
\label{sec:prop}

\subsection{ReLU RNN controller}
To gain full advantage of the non-linearity capability in a data-driven way, we replace the whole FOC controller Eq. (\ref{eq:FOC}) into a RNN. This is a kind of states-to-voltages function that
\begin{align}
\label{eq:rnn}
{\mathbf z} &\equiv \left[ {\hat \omega_e}, \omega_e, i_d, i_q \right]^{\rm T} \notag \\
{\mathbf h}(t+\Delta t) &= \max({\mathbf A} \cdot {\mathbf h}(t)+{\mathbf B} \cdot {\mathbf z}(t)+{\mathbf b_1}, {\mathbf 0}) \notag \\
{\mathbf v}_{dq} &\equiv \left[ v_d, v_q \right]^{\rm T} \notag\\
{\mathbf v}_{dq}(t+\Delta t) &= {\mathbf C} \cdot {\mathbf h}(t+\Delta t) + {\mathbf b_2} \notag \\
 {\mathbf A} &\in {\mathbb R}^{N_h \times N_h}, {\mathbf B} \in {\mathbb R}^{N_h \times 4}, {\mathbf C} \in {\mathbb R}^{2 \times N_h }, \notag \\
 {\mathbf b_1} &\in {\mathbb R}^{N_h}, {\mathbf b_2} \in {\mathbb R}^{2} 
\end{align}
where $N_h$ is the hidden layer size of the RNN, the superscript $\rm T$ means transpose, the Rectified Linear Unit (ReLU\cite{NIPS2000_c8cbd669, pmlr-v15-glorot11a}) function $\max(\cdot,{\mathbf 0})$ is memberwise, and $\Delta t$ is the discretization time width. As we discussed, there is no current references ${\hat i_d}$ or ${\hat i_q}$. Instead of explicitly specifying the current vector, which limits vector control trajectories to the best of our knowledge, we expect the RNN to flexibly walk through the flux-weakening region to maximally improve the transient response.

We select an RNN with an ReLU activation function over other famous RNN models because:
\begin{itemize}
 \item Long Short-Term Memory (LSTM) is considered to have almost comparable performance\cite{weiss2018practical}. However, it has a more complicated architecture that is relatively not suitable for embedded applications.
 \item The Gated Recurrent Unit (GRU) is simpler than LSTM but is known to have theoretical suboptimality\cite{weiss2018practical} compared with the ReLU RNN or LSTM.
\end{itemize}
In terms of computational loads, our controller consists of very simple operations: multiply, add, and sign bit elimination. This property is suitable for time and cost-intensive motor control applications.

\subsection{Lipschitz regularization}
For the purpose of stabilization, we considered regularizing the matrix $\mathbf A$ of the RNN Eq. (\ref{eq:rnn}) because the instability of the controller itself should be avoided. 

We use the decomposition techniques proposed for Lipschitz RNNs \cite{erichson2021lipschitz}
\begin{eqnarray}
 {\mathbf A} &=& (1-\beta) ({\mathbf M}+{\mathbf M}^{\rm T}) + \beta({\mathbf M}-{\mathbf M}^{\rm T}) - \gamma {\mathbf I}
\end{eqnarray}
where $\mathbf I$ is the identity matrix. In this paper, $\mathbf M$ is randomly initialized by an Xavier uniform distribution (also called Glorot initialization)\cite{pmlr-v9-glorot10a} with a gain value of 0.1. $\beta \in [0,1]$ is a hyperparameter that governs the matrix $\mathbf A$ so that it is symmetric, not regularized, or antisymmetric if $\beta$ is 0, 0.5, or 1. $\gamma > 0$ is also a hyperparameter that suppresses the diagonal components of the matrix $\mathbf A$. In this paper, the empirical values $\beta=0.85$ and $\gamma = 0.01$ are used. Note that we omit the linear term in the original paper\cite{erichson2021lipschitz} because of the performance, which may be a consequence of the plant's strong non-linear behavior, and the output of the controller is not in gradient form (which means voltage is directly output) because the plant in the state feedback system is a differential equation that already has the differential operator $\frac{{\rm d}}{{\rm d} t}$.

\subsection{Optimal control-based loss function}
Our purpose is to minimize the speed transient response while making sure that the copper loss is not that large. Thus, we need to define the loss functions over the response trajectories. A similar loss design strategy on PI-FOC can be found in electric bus speed control applications, \cite{9668558}. The straight-forward approach is minimizing the batch mean settling time and copper loss. However, the former cannot be defined because there are the cases that the settling time is not defined when speed diverges or not settled in a fixed simulation time interval $t_{\rm sim}$. Instead of settling time, the area of the speed error in a simulated time interval, which is always defined, can also be used to minimize the settling time. If the area of the speed error is zero, the settling time cannot be improved furthermore. Therefore, assuming discretization, the speed loss is
\begin{align}
{\mathcal L}_s = \frac{1}{N_{\rm batch}} \frac{1}{N_{\rm time}} \sum_{n,t} \frac{|{\hat \omega}_e^{(n)}(t) - \omega_e^{(n)}(t)|}{{\hat \omega}_e^{(n)}(t)}
\end{align}
where $n$ is a sample index in a batch or a dataset, $N_{\rm batch}$ is the number of data, and $N_{\rm time}$ is the number of samples that have the relationship $t_{\rm sim} = N_{\rm time} \cdot \Delta t$. We normalize this quantity to avoid overrating the high speed response. The copper loss is straightforwardly defined as the ratio of the input and the output energy.
\begin{align}
{\mathcal L}_c = \frac{1}{N_{\rm batch}} \sum_{n,t} \frac{R(i_d^{(n)}(t)^2+i_q^{(n)}(t)^2)}{(v_d^{(n)}(t) i_d^{(n)}(t) + v_q^{(n)}(t) i_q^{(n)}(t))} \Delta t
\end{align}
We add overshoot loss to the two losses, which aims at suppressing the amount of overshoot
\begin{align}
{\mathcal L}_o = \frac{1}{N_{\rm batch}} \sum_{n} {\max}_t \max(-\frac{{\hat \omega}_e^{(n)}(t) - \omega_e^{(n)}(t)}{{\hat \omega}_e^{(n)}(t)},0)
\end{align}
and the final value loss, which aims at eliminating steady-state errors.
\begin{align}
{\mathcal L}_f = \frac{1}{N_{\rm batch}} \sum_{n} |\frac{{\hat \omega}_e^{(n)}(N_{\rm time}) - \omega_{e}^{(n)}(N_{\rm time})}{{\hat \omega}_{e}^{(n)}(N_{\rm time})}|
\end{align}
From the optimization issues, we initially train the controller from scratch (random initialization) using only the speed-related loss functions ${\mathcal L}_s + {\mathcal L}_o + {\mathcal L}_f$ for the first 50 epochs (number of parameter updates). After that, we use the full loss functions including copper loss ${\mathcal L}_s + {\mathcal L}_o + {\mathcal L}_f + {\mathcal L}_c$.

\subsection{Circular clamp function in ${\mathbf v}_{dq}$ space}
As PMSMs are typically operated by direct current using inverters, which are easier to handle rather than analog sinusoidal voltage, the maximum voltage $V_{\rm max}$ exists. The rating forms a circle in ${\mathbf v}_{dq}$ space. To limit the output of the NN, we defined a circular clamp function as ${\mathbf v}_{dq} \,\, {\rm if} \,\, \|{\mathbf v}_{dq}\|_2 < V_{\rm max} \,\, {\rm else}\,\, V_{\rm max}$.

\section{Simulation}
\label{sec:exp}
\subsection{Plant}
The target IPMSM plant was an IEEJ D1\cite{IEE1296}-like model, shown in Table \ref{table:D1}. 
\subsection{Data}
\begin{figure}[t]
 \begin{tabular}{c}
 \begin{minipage}{0.45\hsize}
 \centering
 \includegraphics[width=4.2cm]{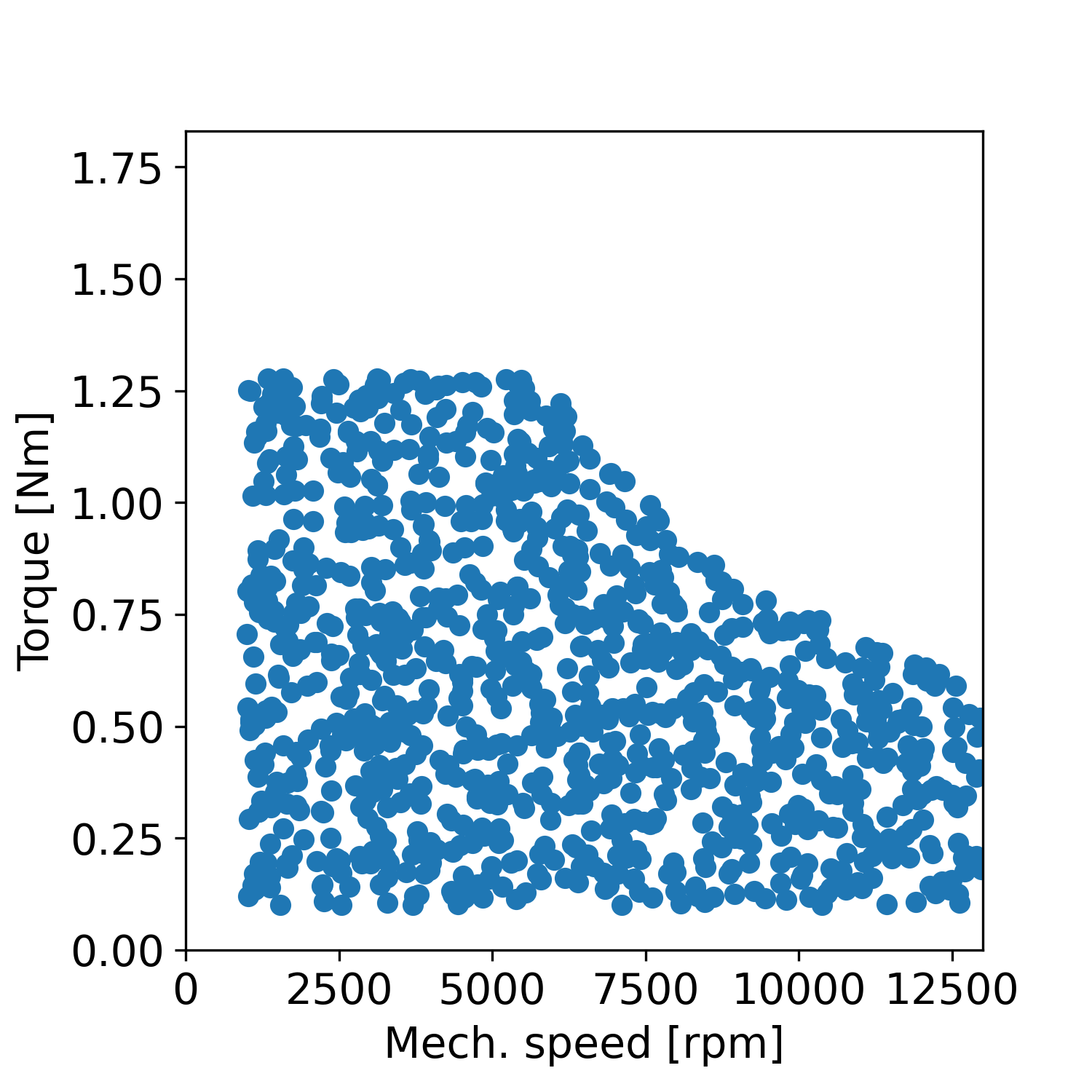}
 \centerline{(a) Training set}\medskip
 \end{minipage}
 \begin{minipage}{0.45\hsize}
 \centering
 \includegraphics[width=4.2cm]{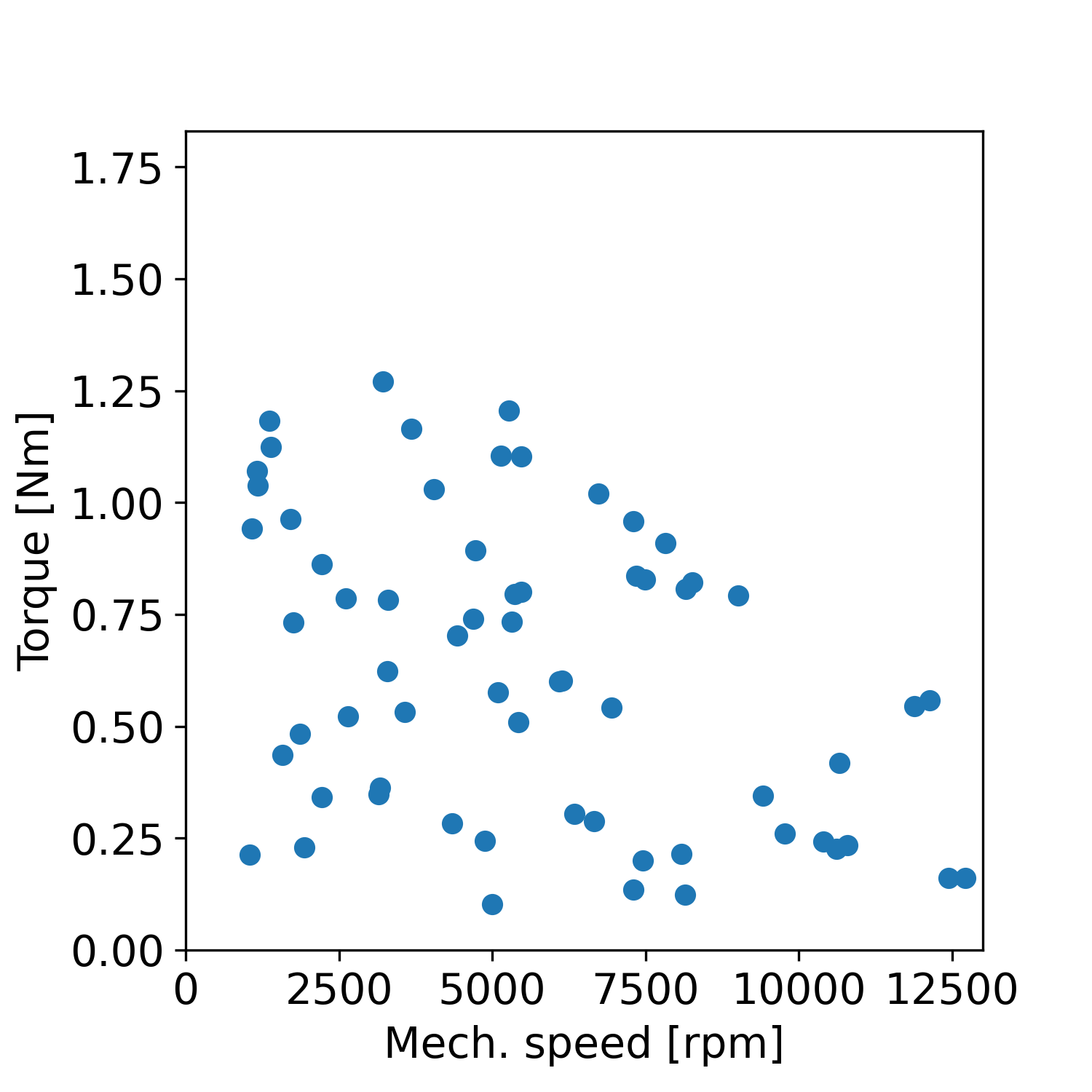} 
 \centerline{(b) Evaluation set}\medskip
 \end{minipage}
 \end{tabular}
\caption{Dataset definition. $N_{\rm batch}$ points in training dataset are sampled per epoch.}
\label{fig:dataset}
\end{figure}

Unlike the typical ML settings, we arranged the data artificially. The data were arranged in a speed-torque space inside the speed and torque range. As the mechanical power is $\omega_m \cdot T_L$, where $P\omega_m = \omega_e$, the samples over the maximum value $\omega_m \cdot T_L > P_{\rm max}$ were also removed to avoid impractical settings. The training and evaluation data was randomly generated in this region, which is shown in the Fig. \ref{fig:dataset}. Having these as the final values, we specified the acceleration time $t_{\rm ramp}=1.0\,s$ to be common instead of the ramp slope to generate saturated-ramp speed reference functions.

\subsection{Simulator}
As the state feedback system with plant is also an ODE, the whole system was simulated using the classical Runge-Kutta method (known as RK4) unrolled in a computational graph. $\Delta t=2 \times 10^{-4}\,s$, and $t_{\rm sim} = 2\,s$.

\subsection{Initial states}
Basically, zeros are used. But for training the RNN, the zero-centered uniform distributions with $i_d, i_q \in [-2.5, 2.5)\,A$, $\omega_e \in [ -100P\frac{2\pi}{60},100P\frac{2\pi}{60})\rm\,rad/s$ were used in order to increase the robustness.

\subsection{Optimization techniques}
Instead of the classical SGD, Adam\cite{kingma2015Adam} was used because it is empirically good at training difficult tasks, such as time-dependent tasks. The minibatch size $N_{\rm batch}$ was 8. As NNs are empirically not good at dealing with values larger than one, the ${\mathbf v}_{\rm dq}$ values from the controllers were multiplied by $V_{\rm max}$ for normalization. The whole system was implemented in PyTorch.

\subsection{Conventional PI-FOC}
To maximally widen the supported speed range, we set the baseline as the MC controller. The controller just tried to keep the current vector on the maximum current circle with the radius $I_{\rm max}$ using ${\hat i_q}$ from the speed controller. The gains are {listed} in Table \ref{table:gains}.

\subsection{Proposed RNN}
The hidden layer size of the RNN $N_{h}$ was 128. $\mathbf B$ was initialized using the Xavier uniform distribution with a gain of $1\times 10^{-6}$, $\mathbf C$ was initialized using a uniform distribution with $\pm 1 \times 10^{-6}$ as this matrix should empirically be started from near-zero, and the initial bias vectors were ${\mathbf b}_1={\mathbf b}_2={\mathbf 0}$.

\begin{table}[tb]
\caption{Simulated IPMSM parameters}
\label{table:D1}
\begin{center}
\begin{tabular}{c|c|c}
\noalign{\hrule height 0.4mm}
 symbol & value & unit\\
\hline
$R$ & 0.38 & [$\rm \Omega$] \\
$L_d$ & $11.2 \times 10^{-3}$ & [H] \\
$L_q$ & $19 \times 10^{-3}$ & [H] \\
${\rm \Phi}$ & 0.107 & [Wb] \\
$P$ & 2 & [] \\
$J$ & $10 \times 10^{-4}$ & [$\rm kgm^2$] \\
$D$ & 0 & [$\rm Nm s/rad$] \\
$V_{\rm max}$ & 233 & [V] \\
$I_{\rm max}$ & 13 & [A] \\
$P_{\rm max}$ & 800 & [W] \\
$f_{\rm min}$ & 1000 & [rpm] \\
$f_{\rm max}$ & 13000 & [rpm] \\
$T_{L{\rm min}}$ & 0.1 & [Nm] \\
$T_{L{\rm max}}$ & 1.83 & [Nm] \\
\noalign{\hrule height 0.4mm}
\end{tabular}
\end{center}
\end{table}

\begin{table}[tb]
\caption{PI-FOC gain values, where integral time $T_{i*}={k_{p*}}/{k_{i*}}$ was tuned almost comparatively under acceleration time 1 s used to train RNN. Note that the extra maximum and minimum values (as anti-windup or current reference limiters) are introduced only on the faster speed ramp condition with the acceleration time of 0.2 s to support high-speed operation area, which is not covered without introducing these limiters.}
\label{table:gains}
\begin{center}
\begin{tabular}{c|c|c}
\noalign{\hrule height 0.4mm}
 meaning & symbol & value \\
\hline

 speed proportional gain &  $k_{p\omega}$ & 0.100 \\
 speed integral time  &  $T_{i\omega}$ & 0.100 \\
 d-axis current proportional gain &  $k_{pd}$ & 5.60 \\
 d-axis current integral time &  $T_{id}$ & 0.0295 \\
 q-axis current proportional gain &  $k_{pq}$ & 9.50 \\
 q-axis current integral time &  $T_{iq}$ & 0.0500\\\hline
 elec. speed anti-windup (max) & $s_{\omega{\rm max}}$ & 5 \\
 elec. speed anti-windup (min) & $s_{\omega{\rm min}}$ & -1 \\
 d-axis anti-windup value (max) &  $s_{id{\rm max}}$ & 1 \\
 d-axis anti-windup value (min) &  $s_{id{\rm min}}$ & -0.03 \\
 q-axis anti-windup value (max) &  $s_{iq{\rm max}}$ & 0.02 \\
 q-axis anti-windup value (min) &  $s_{iq{\rm min}}$ & -0.01 \\
 d-axis current reference (max)  &  ${\hat i_{d{\rm max}}}$ & -5 \\
 d-axis current reference (min)  &  ${\hat i_{d{\rm min}}}$ & -100 \\
 q-axis current reference (max) &  ${\hat i_{q{\rm max}}}$ & 8 \\
 q-axis current reference (min) &  ${\hat i_{q{\rm min}}}$ & -100 \\
\noalign{\hrule height 0.4mm}
\end{tabular}
\end{center}
\end{table}

\subsection{Results}
\begin{figure}[t]
\begin{center}
 \includegraphics[width=\linewidth]{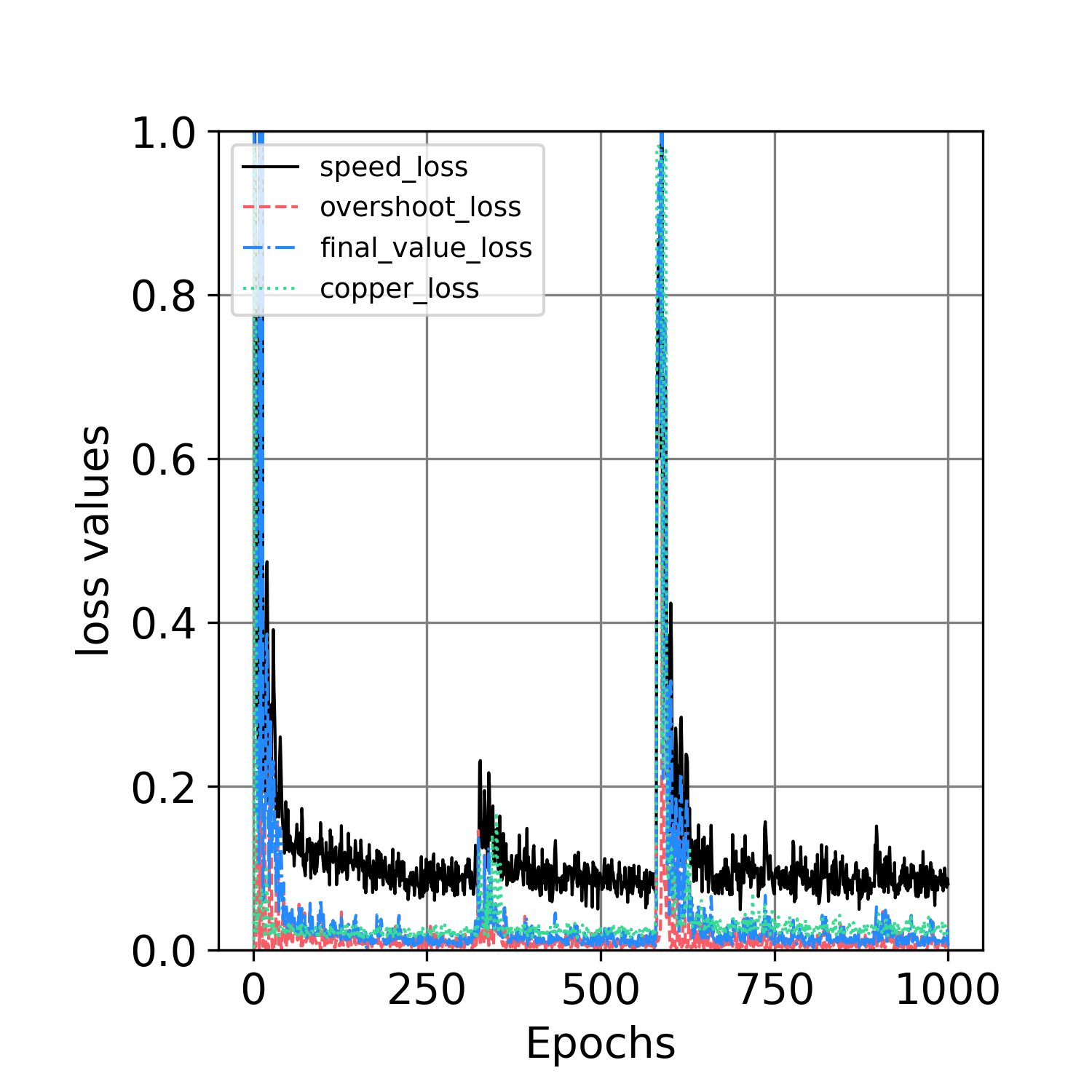}
\end{center}
\caption{Training loss throughout RNN training epoch}
\label{fig:loss_train}
\end{figure}

\begin{figure}[t]
\begin{center}
 \includegraphics[width=\linewidth]{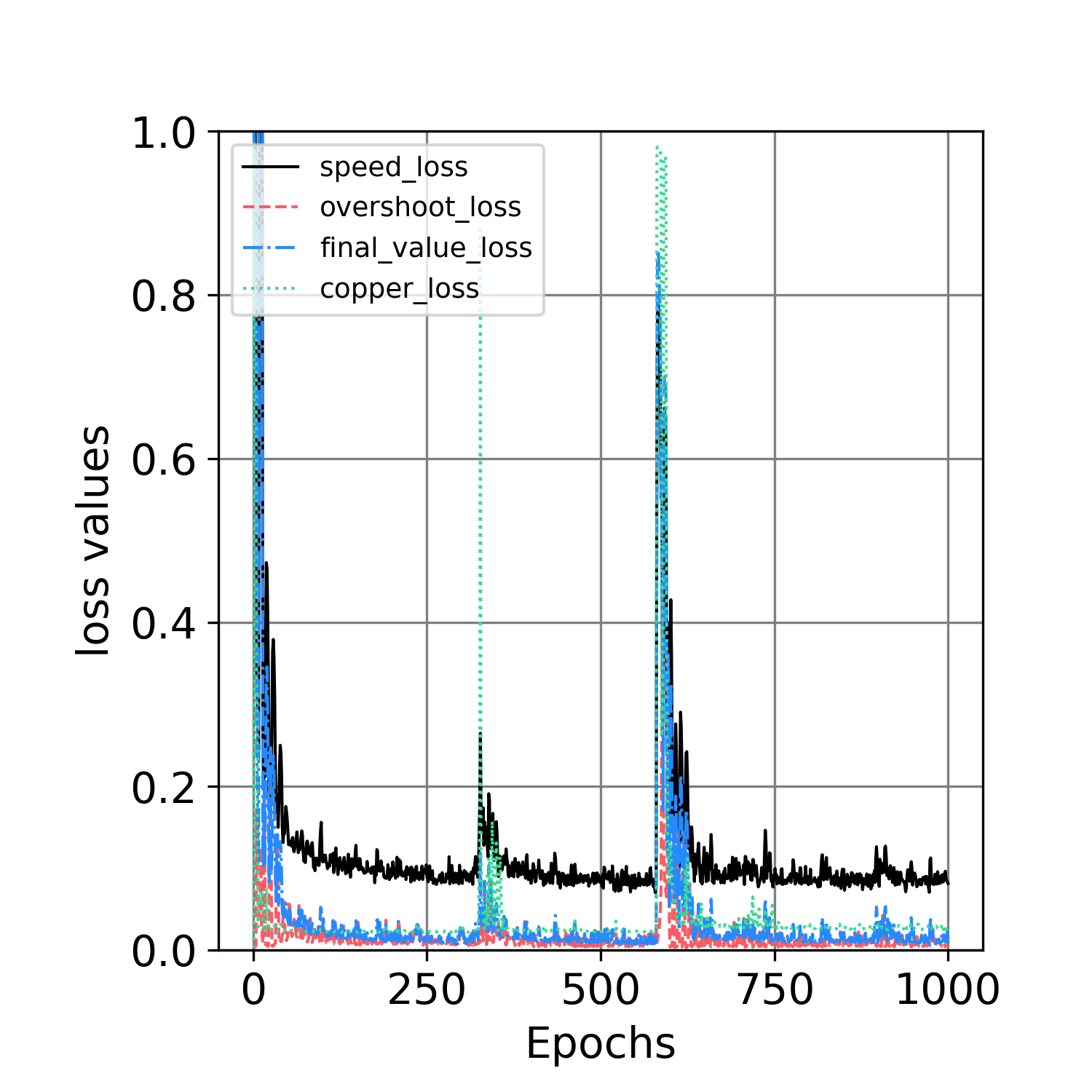}
\end{center}
\caption{Evaluation loss throughout RNN training epoch}
\label{fig:loss_dev}
\end{figure}

\begin{figure}[t]
\begin{center}
 \includegraphics[width=\linewidth]{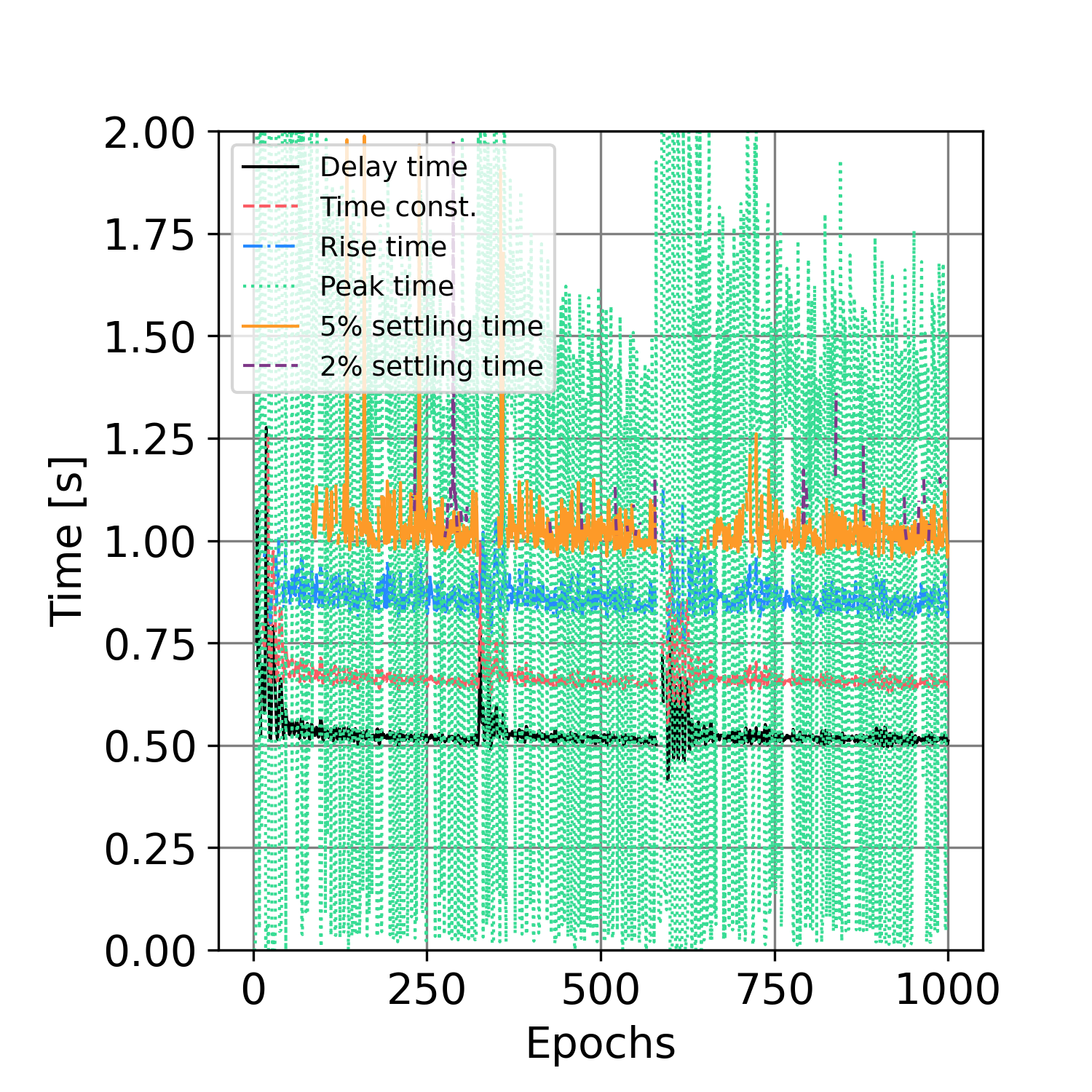}
\end{center}
\caption{Training response indices throughout RNN training epoch}
\label{fig:indices_train}
\end{figure}

\begin{figure}[t]
\begin{center}
 \includegraphics[width=\linewidth]{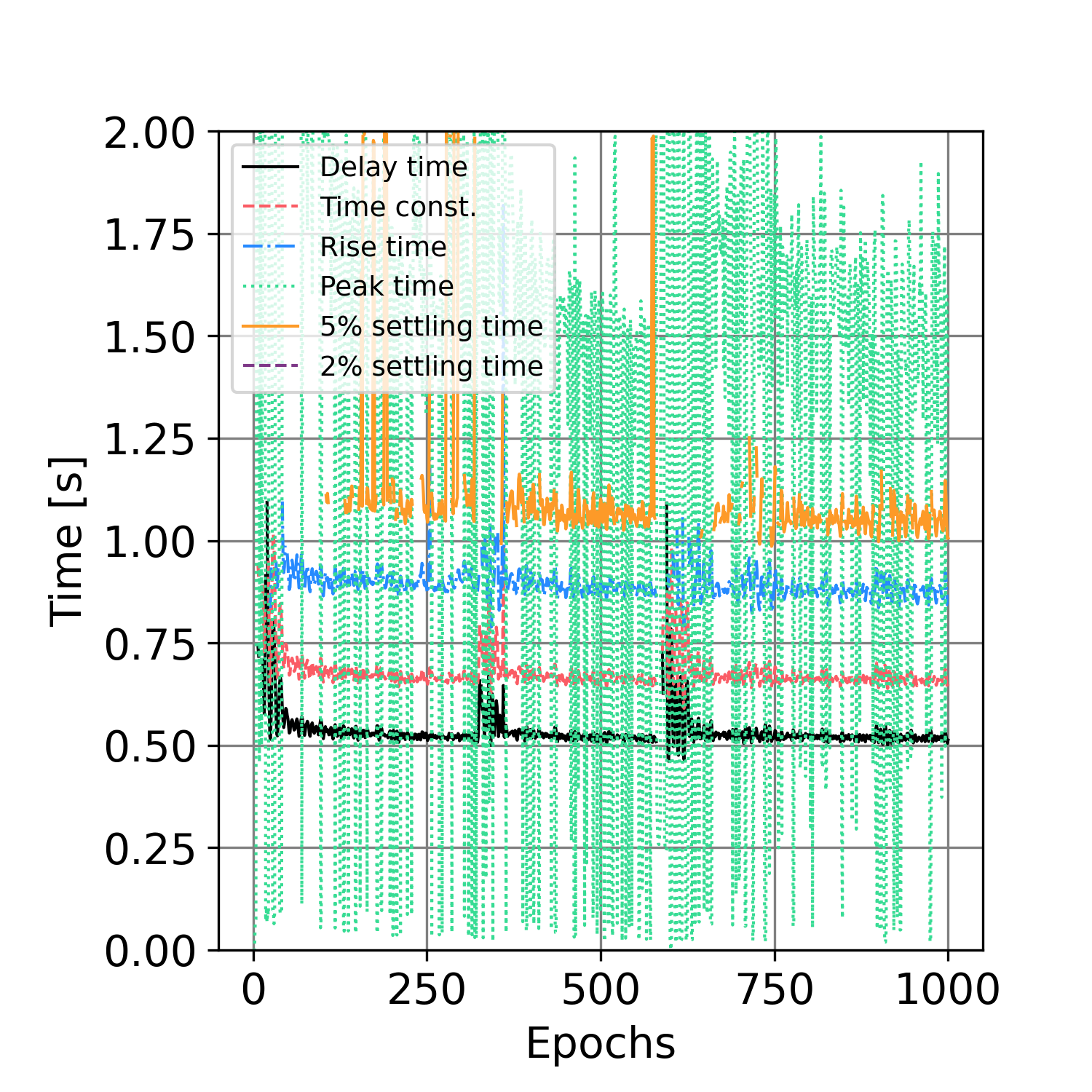}
\end{center}
\caption{Evaluation response indices throughout RNN training epoch}
\label{fig:indices_dev}
\end{figure}

\begin{figure}[t]
\begin{center}

\includegraphics[width=\linewidth]{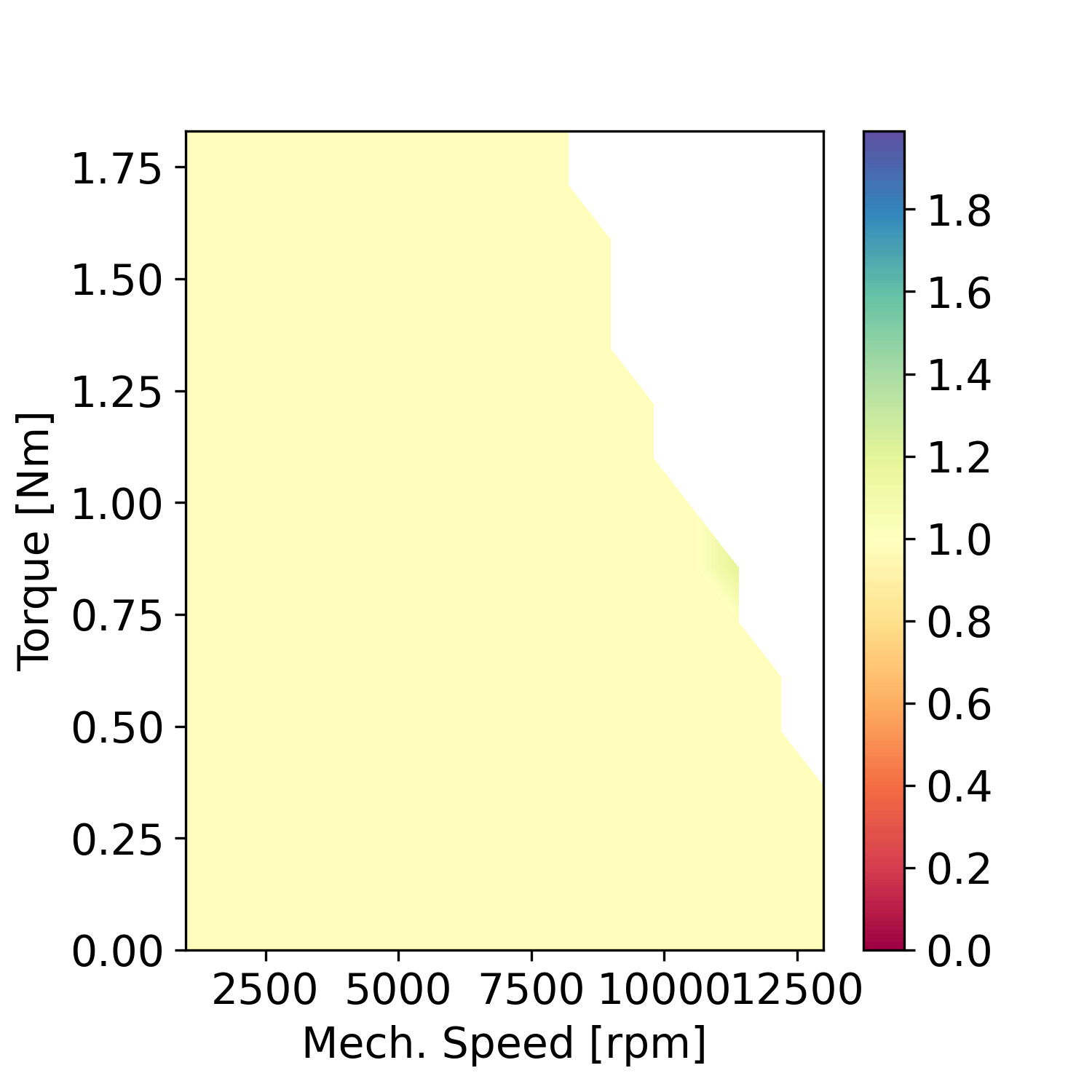}
\end{center}
\caption{2\% settling time [s] using PI-FOC (MC) under 1.0 s acceleration time.}
\label{fig:stt2_devel_mc}
\end{figure}

\begin{figure}[t]
\begin{center}

 \includegraphics[width=\linewidth]{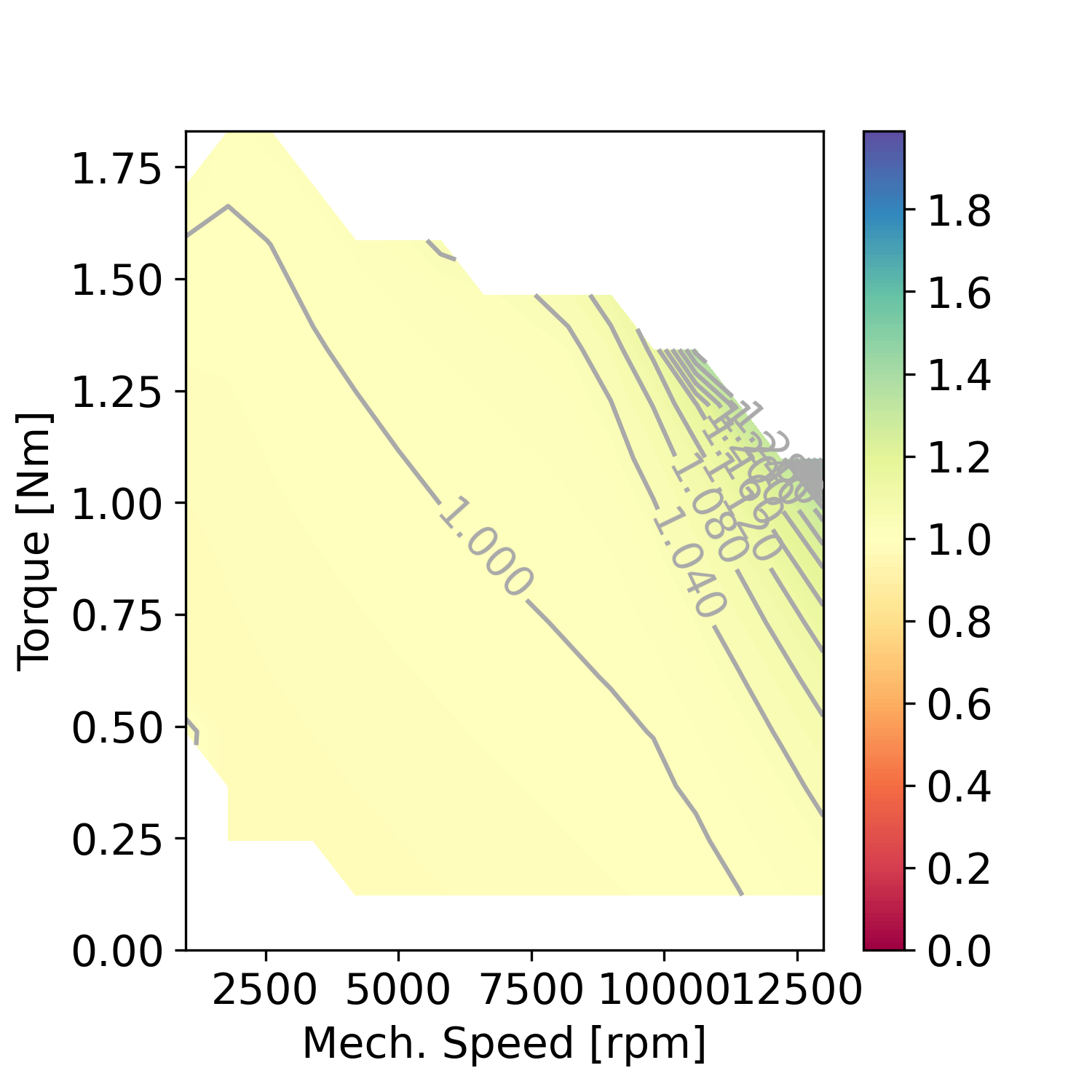}
\end{center}
\caption{2\% settling time [s] using RNN under 1.0 s acceleration time.}
\label{fig:stt2_devel_rnn}
\end{figure}

\begin{figure}[t]
\begin{center}
 \includegraphics[width=\linewidth]{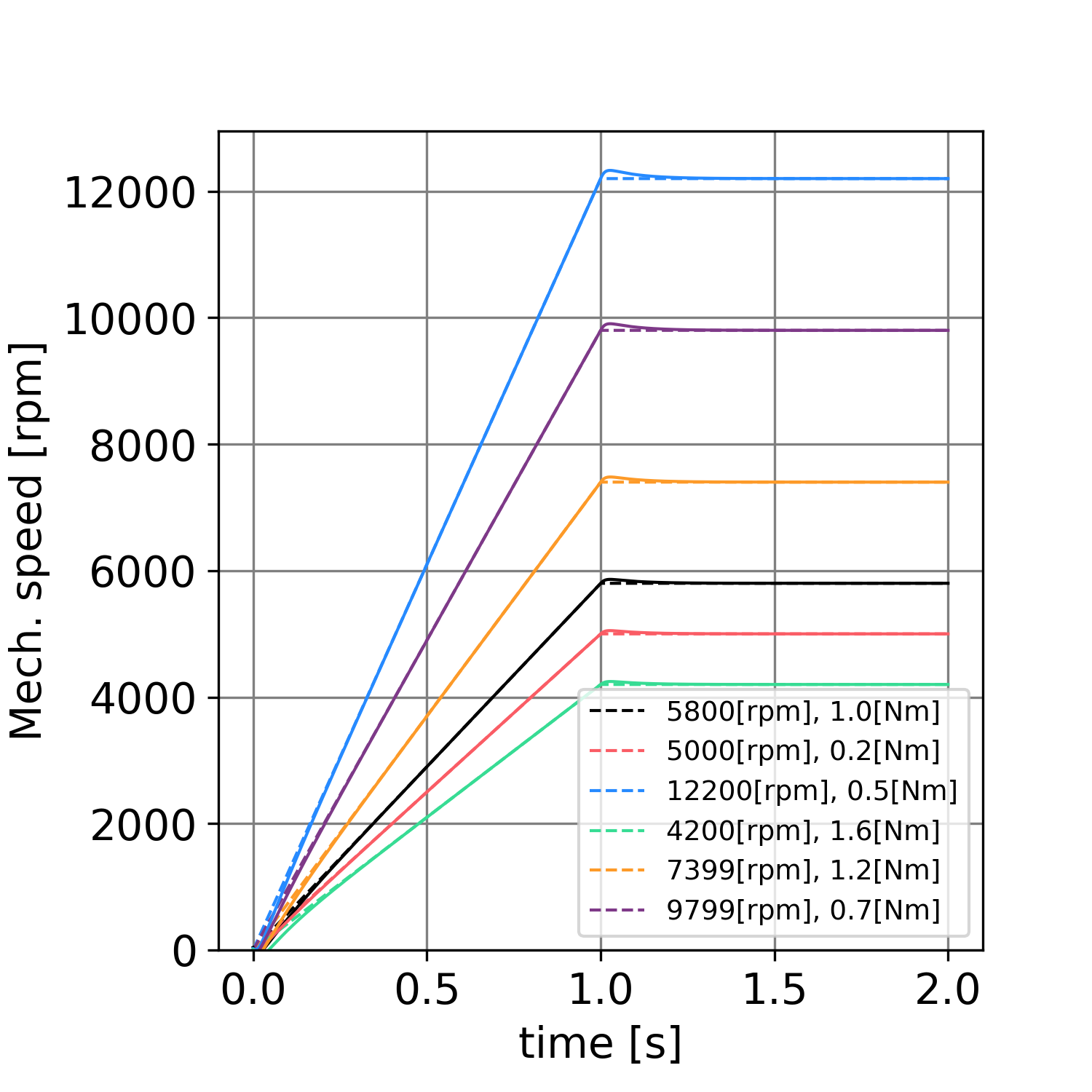}
\end{center}
\caption{Speed response using PI-FOC (MC) under 1.0 s acceleration time. References shown in dotted lines.}
\label{fig:fmfr_devel_mc}
\end{figure}

\begin{figure}[t]
\begin{center}
 \includegraphics[width=\linewidth]{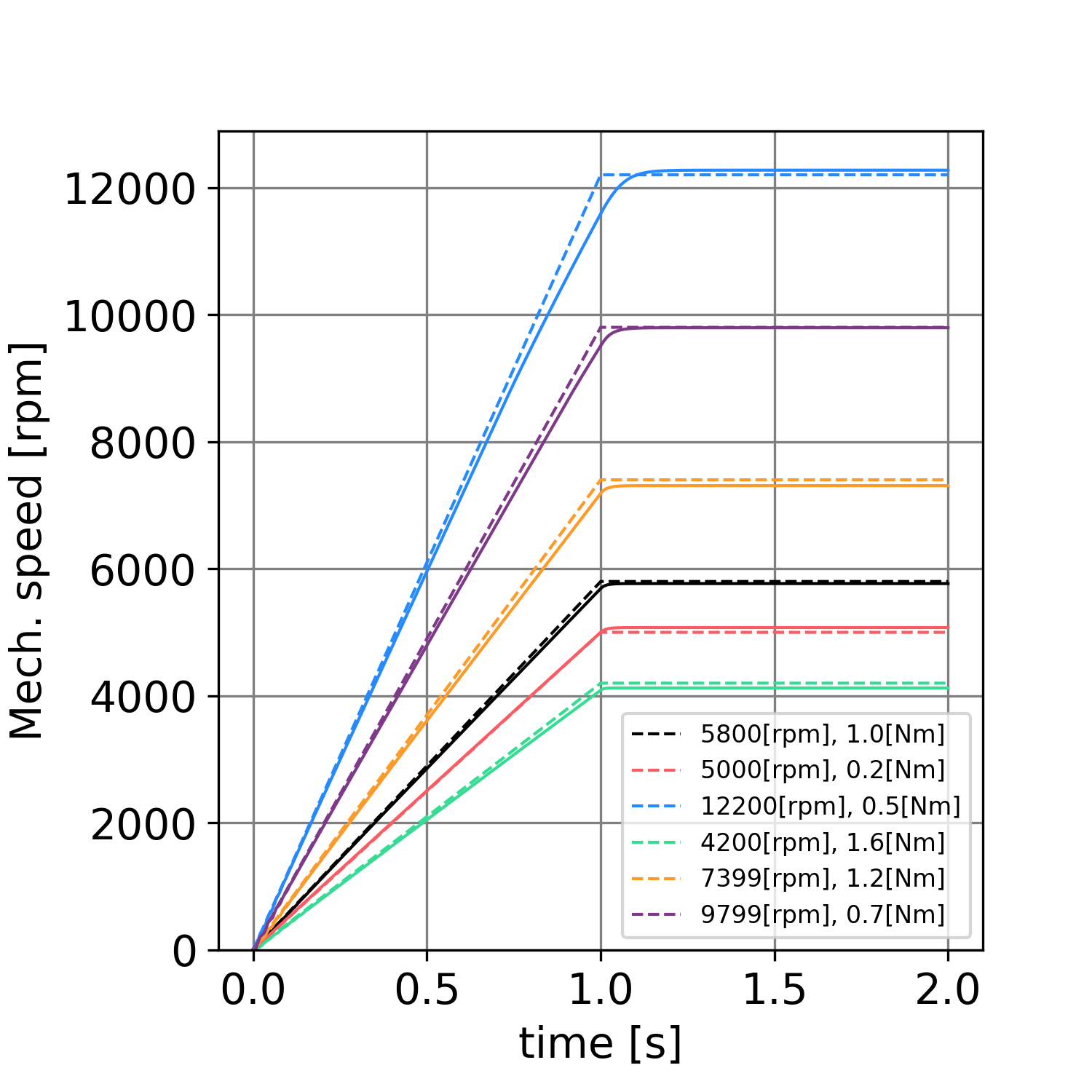}
\end{center}
\caption{Speed response using RNN under 1.0 s acceleration time. References shown in dotted lines.}
\label{fig:fmfr_devel_rnn}
\end{figure}

\begin{figure}[t]
\begin{center}
 \includegraphics[width=\linewidth]{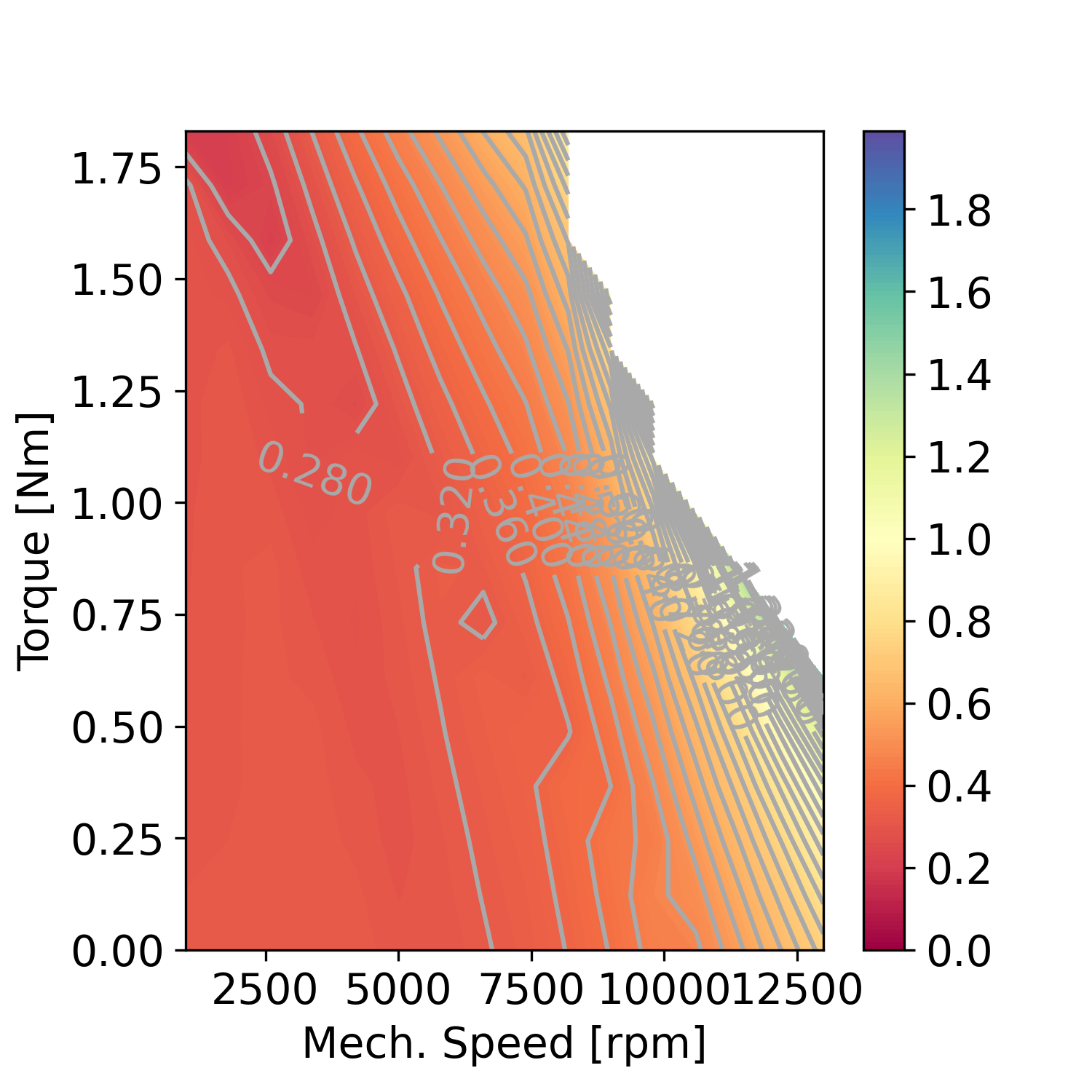}
\end{center}
\caption{2\% settling time [s] using PI-FOC (MC with limiters) under 0.2 s acceleration time.}
\label{fig:stt2_ext_mc}
\end{figure}

\begin{figure}[t]
\begin{center}
 \includegraphics[width=\linewidth]{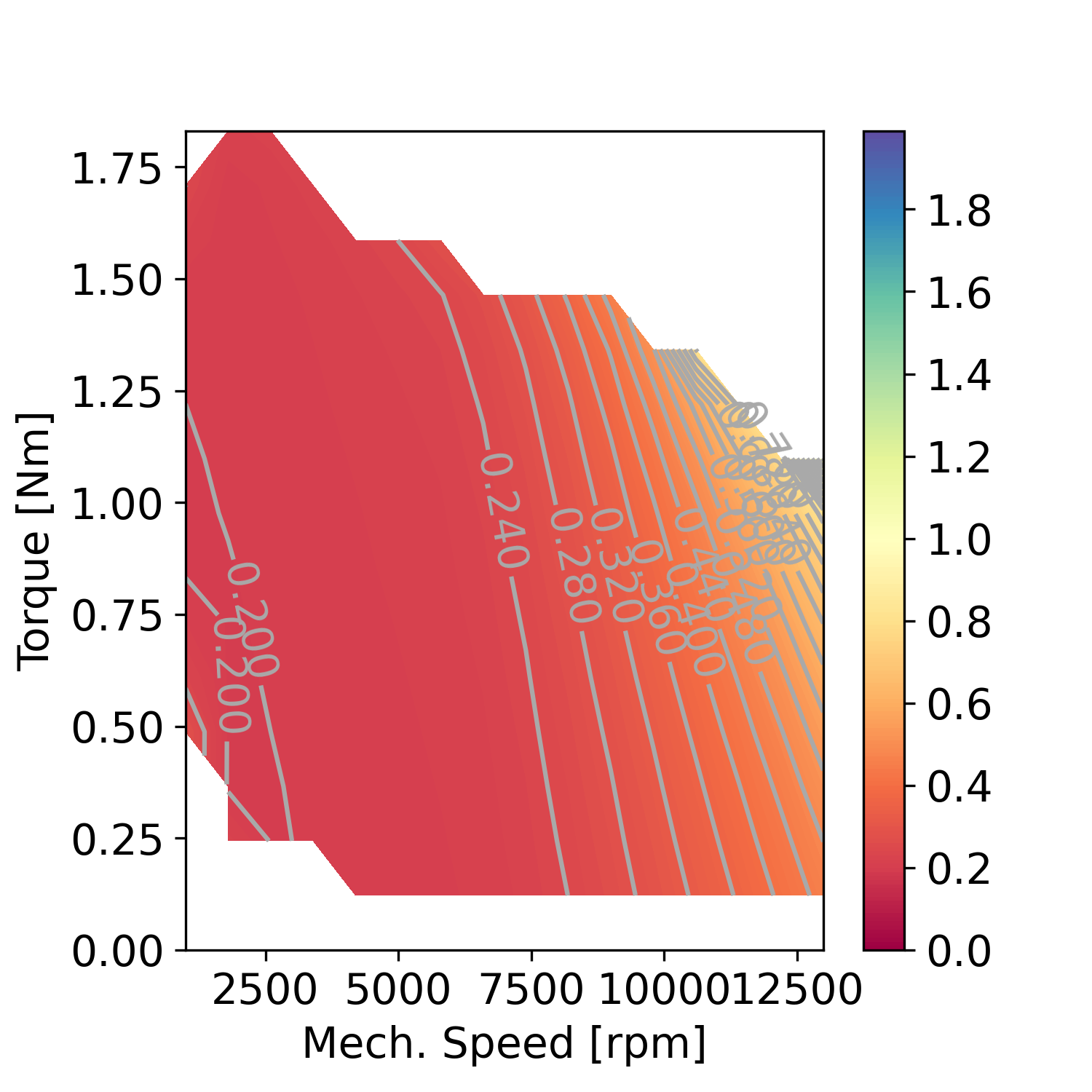}
\end{center}
\caption{2\% settling time [s] using RNN under 0.2 s acceleration time.}
\label{fig:stt2_ext_rnn}
\end{figure}

\begin{figure}[t]
\begin{center}
 \includegraphics[width=\linewidth]{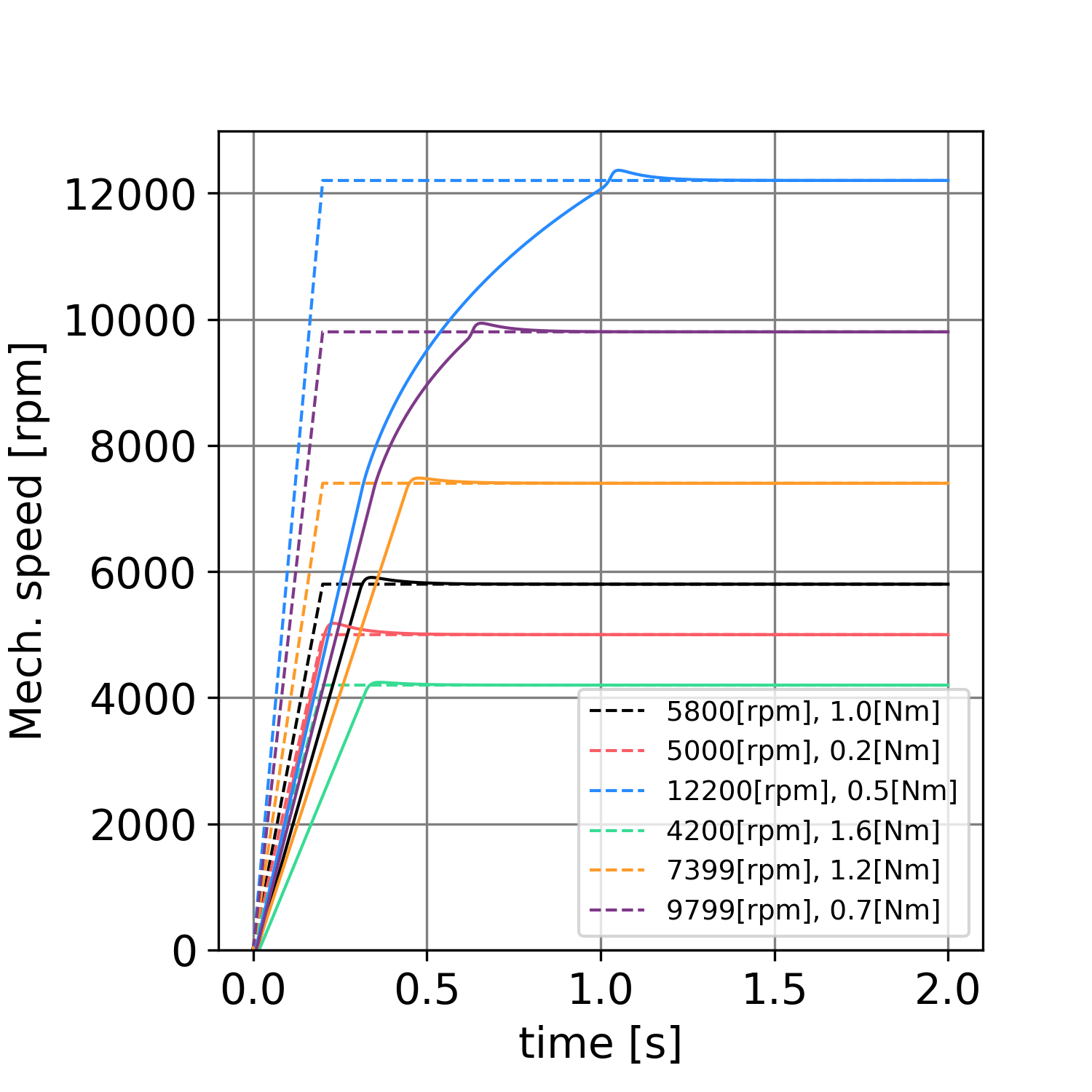}
\end{center}
\caption{Speed response using PI-FOC (MC with limiters) under 0.2 s acceleration time. References shown in dotted lines.}
\label{fig:fmfr_ext_mc}
\end{figure}

\begin{figure}[t]
\begin{center}
 \includegraphics[width=\linewidth]{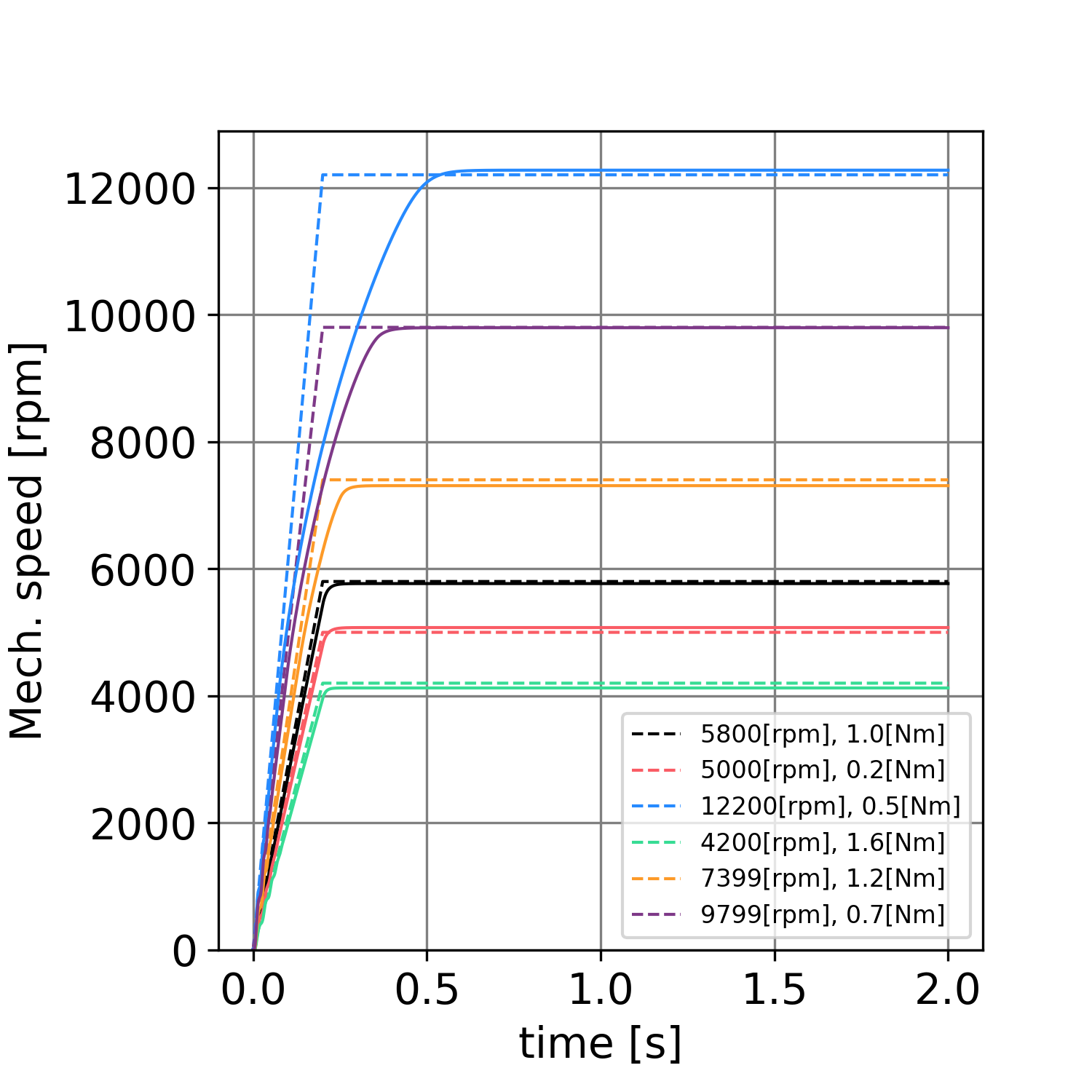}
\end{center}
\caption{Speed response using RNN under 0.2 s acceleration time. References shown in dotted lines.}
\label{fig:fmfr_ext_rnn}
\end{figure}

\begin{figure}[t]
\begin{center}
  \includegraphics[width=\linewidth]{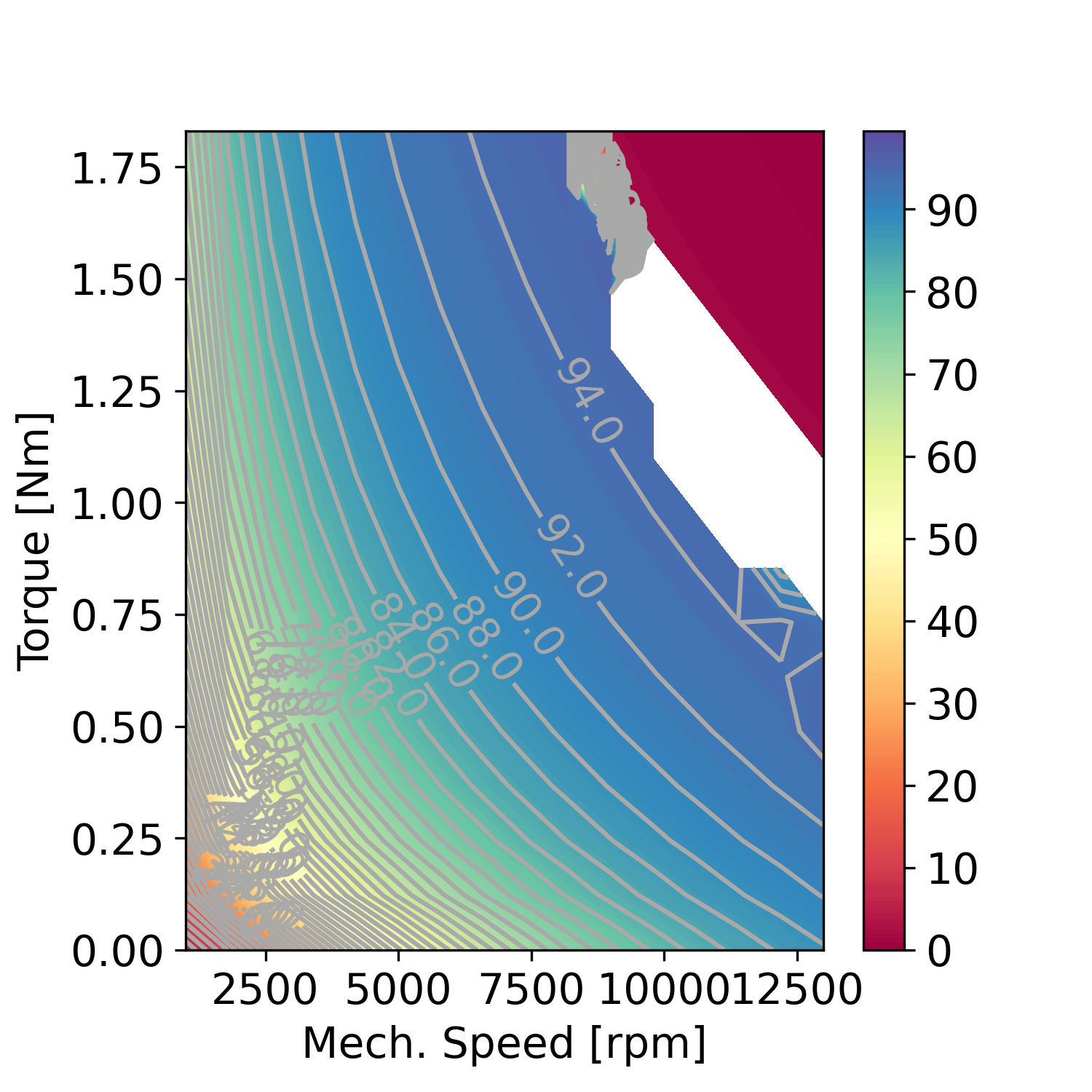}
 
\end{center}
\caption{Motor efficiency [\%] using PI-FOC (MC) under 1.0 s acceleration time.}
\label{fig:eff_devel_mc}
\end{figure}

\begin{figure}[t]
\begin{center}
 \includegraphics[width=\linewidth]{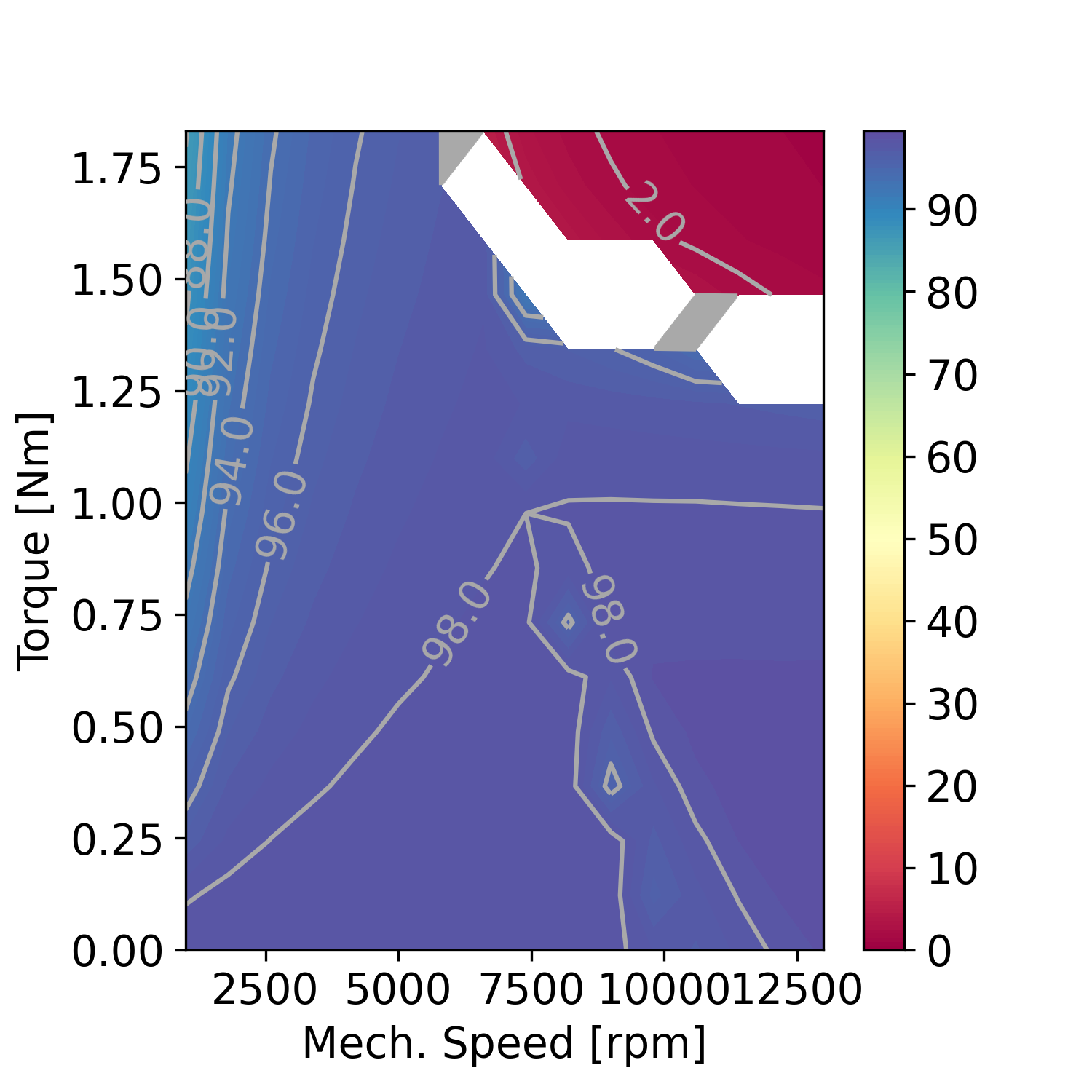}
\end{center}
\caption{Motor efficiency [\%] using PI-FOC (MTPA) under 1.0 s acceleration time.}
\label{fig:eff_devel_mtpa}
\end{figure}

\begin{figure}[t]
\begin{center}
 \includegraphics[width=\linewidth]{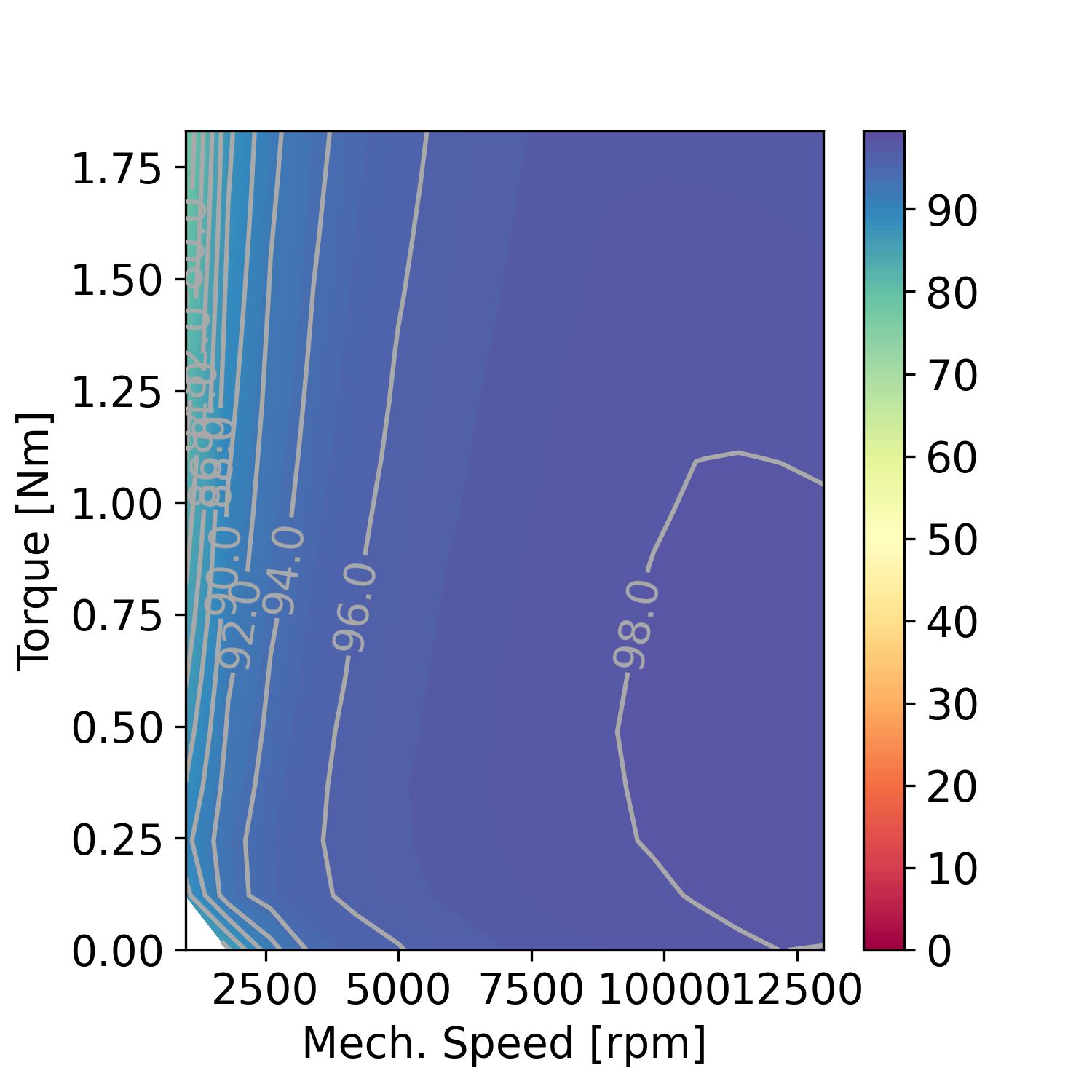}
\end{center}
\caption{Motor efficiency [\%] using RNN under 1.0 s acceleration time.}
\label{fig:eff_devel_rnn}
\end{figure}

\begin{figure}[t]
\begin{center}
 \includegraphics[width=\linewidth]{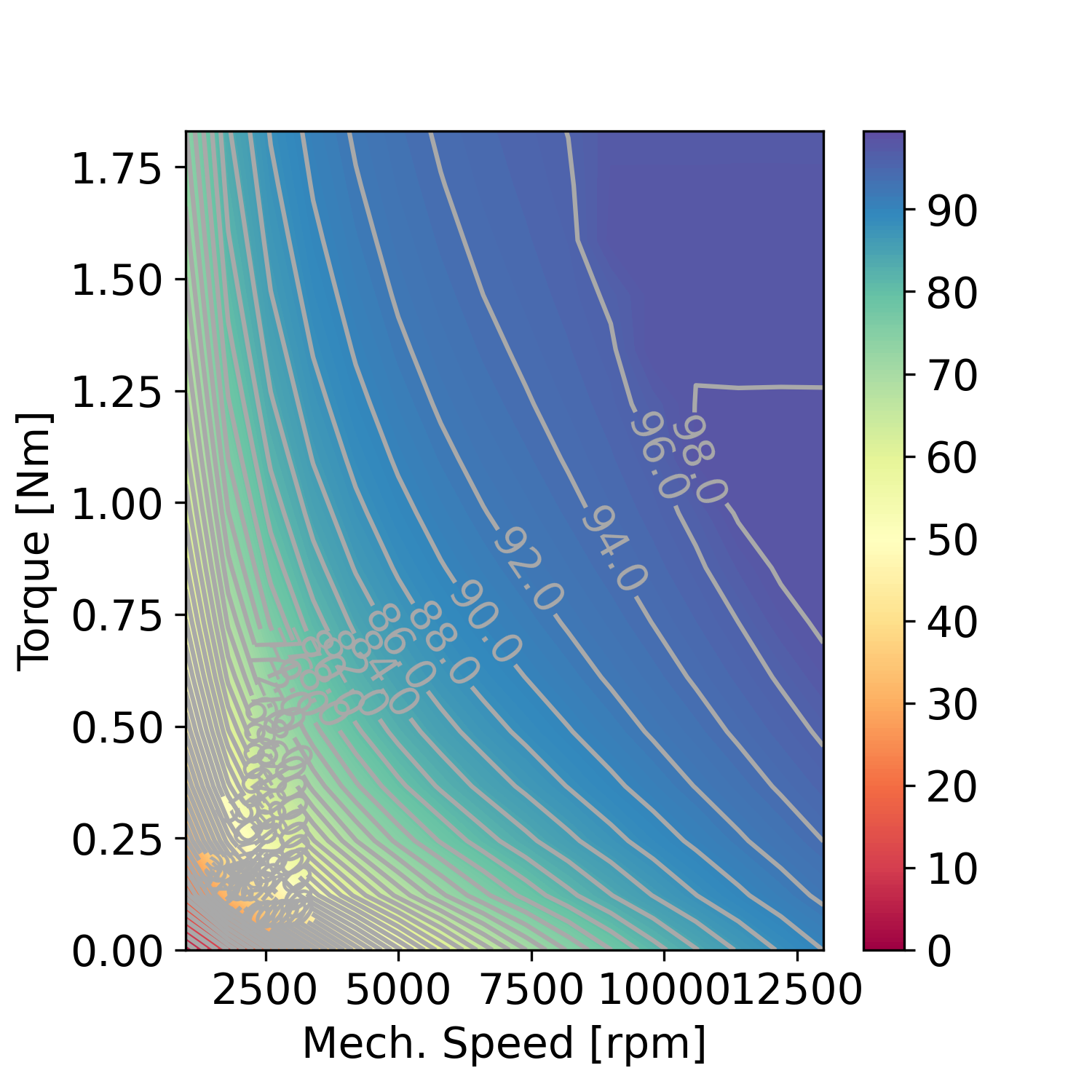}
\end{center}
\caption{Motor efficiency [\%] using PI-FOC (MC with limiters) under 0.2 s acceleration time.}
\label{fig:eff_ext_mc}
\end{figure}

\begin{figure}[t]
\begin{center}
 \includegraphics[width=\linewidth]{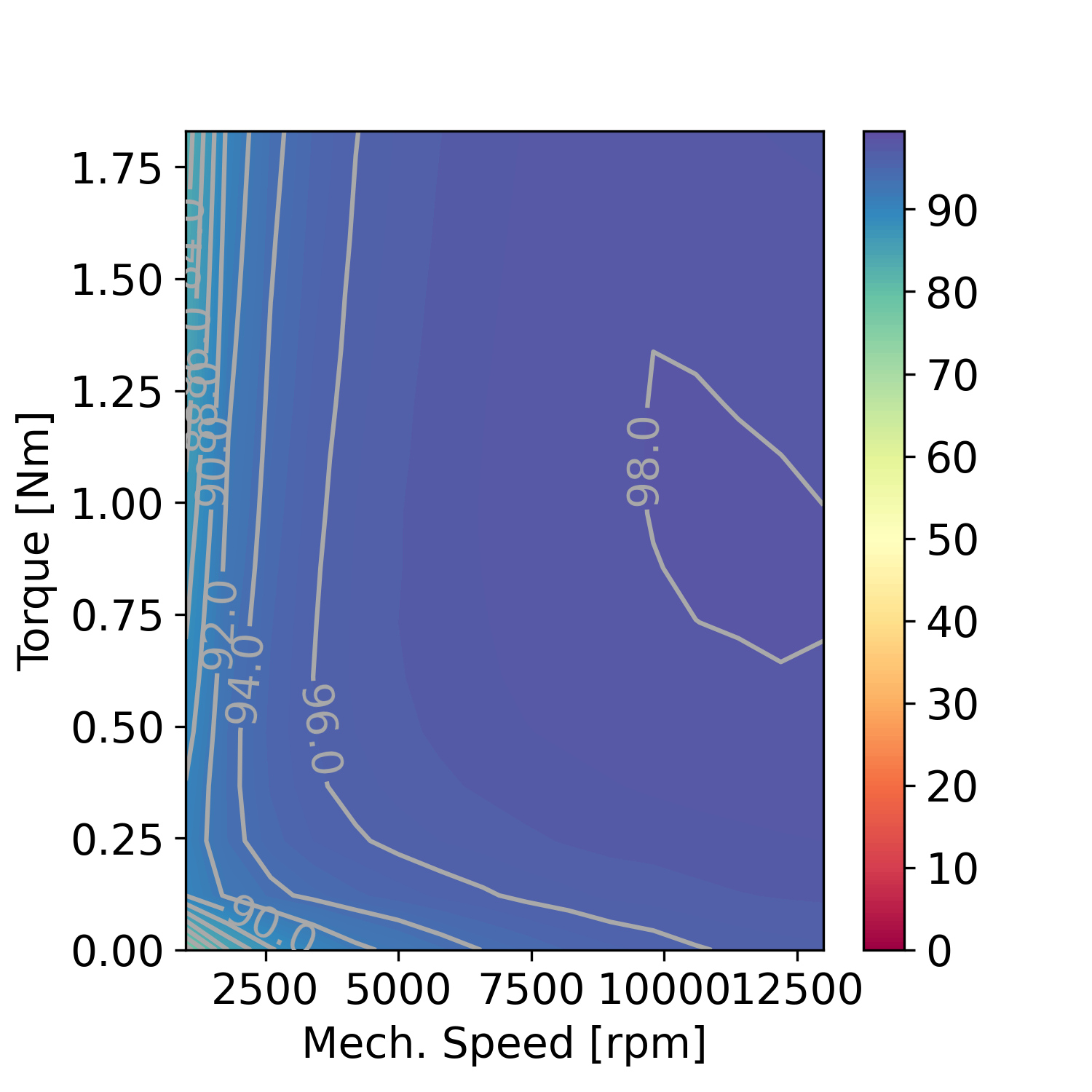}
\end{center}
\caption{Motor efficiency [\%] using RNN under 0.2 s acceleration time.}
\label{fig:eff_ext_rnn}
\end{figure}

The error throughout the epochs on the training set is shown in Fig. \ref{fig:loss_train}.
The error throughout the epochs on the evaluation set is shown in Fig. \ref{fig:loss_dev}. In terms of ML, there is no problematic overfitting since little difference can be found between the training and evaluation error. We also measured the response indices on the training set throughout the epochs, which is shown in Fig. \ref{fig:indices_train}, and the response indices on the evaluation set throughout the epochs, which is shown in Fig. \ref{fig:indices_dev}. From the two figures, the response indices were also showing no problematic overfitting.
We compared 1000 epoch training results of the RNN and the baseline controllers. 
The 2\% settling time using the baseline PI-FOC (MC) controller in the speed-torque plane is shown in Fig. \ref{fig:stt2_devel_mc}, and the 2\% settling time using the proposed RNN controller in the speed-torque plane is shown in Fig. \ref{fig:stt2_devel_rnn}. 
\footnote{Note that only valid (which means the existence of settling times or energy conservation numerically holds due to the implementation) areas are shown.}
The 2\% settling time of two controllers show that the the two controllers were almost comparative in the meaning of the operable area in the speed-torque plane. We also compared the speed transient response on the representative points in the plane. The result of the PI-FOC (MC) controller is shown in Fig. \ref{fig:fmfr_devel_mc} and the result of the RNN controller is shown in Fig. \ref{fig:fmfr_devel_rnn}. From the viewpoint of the overshoot, the two results are also almost comparative.
In most cases, the final values of the RNN controller's response seemed to be at one of the equilibria\footnote{ Note that we roughly compared the final states $(i_d, i_q, \omega_e)$ of the trained RNN and the equilibria numerically calculated by the Newton method using the final $v_{dq}$ values. We observed that the differences were so small that we could say that most of the outputs of the RNN controller finally reached one of the equilibria.}.

To examine the extrapolation to a quicker response, we fed the two controllers a faster ramp with an acceleration time of 0.2 s.

The 2\% settling time of the PI-FOC (MC with limiters) controller in the speed-torque plane is shown in Fig. \ref{fig:stt2_ext_mc} and the 2\% settling time of the RNN controller in the speed-torque plane is shown in Fig. \ref{fig:stt2_ext_rnn}. 
The results show the advantage of the proposed RNN controller, which consistently has a shorter settling time in the large part of the speed-torque plane. From the speed transient response of the PI-FOC (MC with limiters), shown in Fig. \ref{fig:fmfr_ext_mc} and the response of the RNN, shown in Fig. \ref{fig:fmfr_ext_rnn} , we can check that the speed responses of the RNN is substantially faster than PI-FOC. The RNN is especially effective in settling time on high-speed commands. The form of high-speed responses is similar to saturated-exponential functions in the case of PI-FOC and the form in RNN looks like in the middle of the saturated-exponential and saturated-step functions. This supports the hypothesis in Sect \ref{sec:ramp} that the saturated-exponential function is not enough in the meaning of the IPMSM transient response optimization.

In order to evaluate the copper losses of the controllers, we set the baseline copper losses of the simulations of 1.0 s acceleration time. The PI-FOC (MC) baseline is shown in Fig. \ref{fig:eff_devel_mc}, the PI-FOC (MTPA) baseline is shown in Fig. \ref{fig:eff_devel_mtpa}, and the RNN loss is shown in Fig. \ref{fig:eff_devel_rnn}. MTPA is added because it is considered to have near minimum copper loss. In the 0.2 s acceleration time simulations, the copper loss of the PI-FOC (MC with limiters) is shown in Fig. \ref{fig:eff_ext_mc} and the copper loss of the RNN is shown in Fig. \ref{fig:eff_ext_rnn}. From these results, PI-FOC (MC with or without limiters) methods have generally worse copper loss than the PI-FOC (MTPA) or the RNN and the two methods have almost similar copper loss, which is considered to be almost minimum.

\begin{figure}[t]
\begin{center}
 \includegraphics[width=\linewidth]{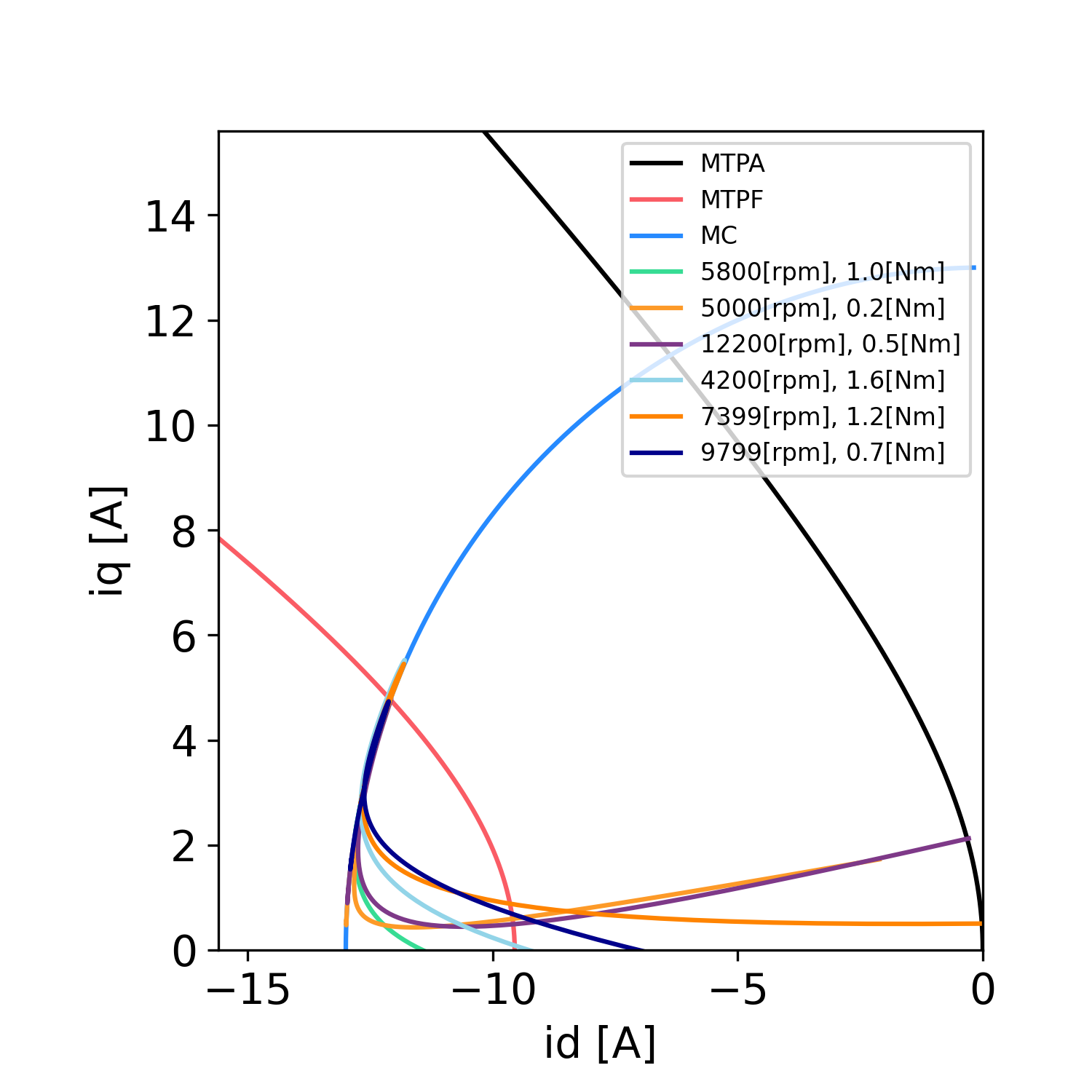}
\end{center}
\caption{Current vector space using PI-FOC (MC) under 1.0 s acceleration time.}
\label{fig:vec_devel_mc}
\end{figure}

\begin{figure}[t]
\begin{center}
\includegraphics[width=\linewidth]{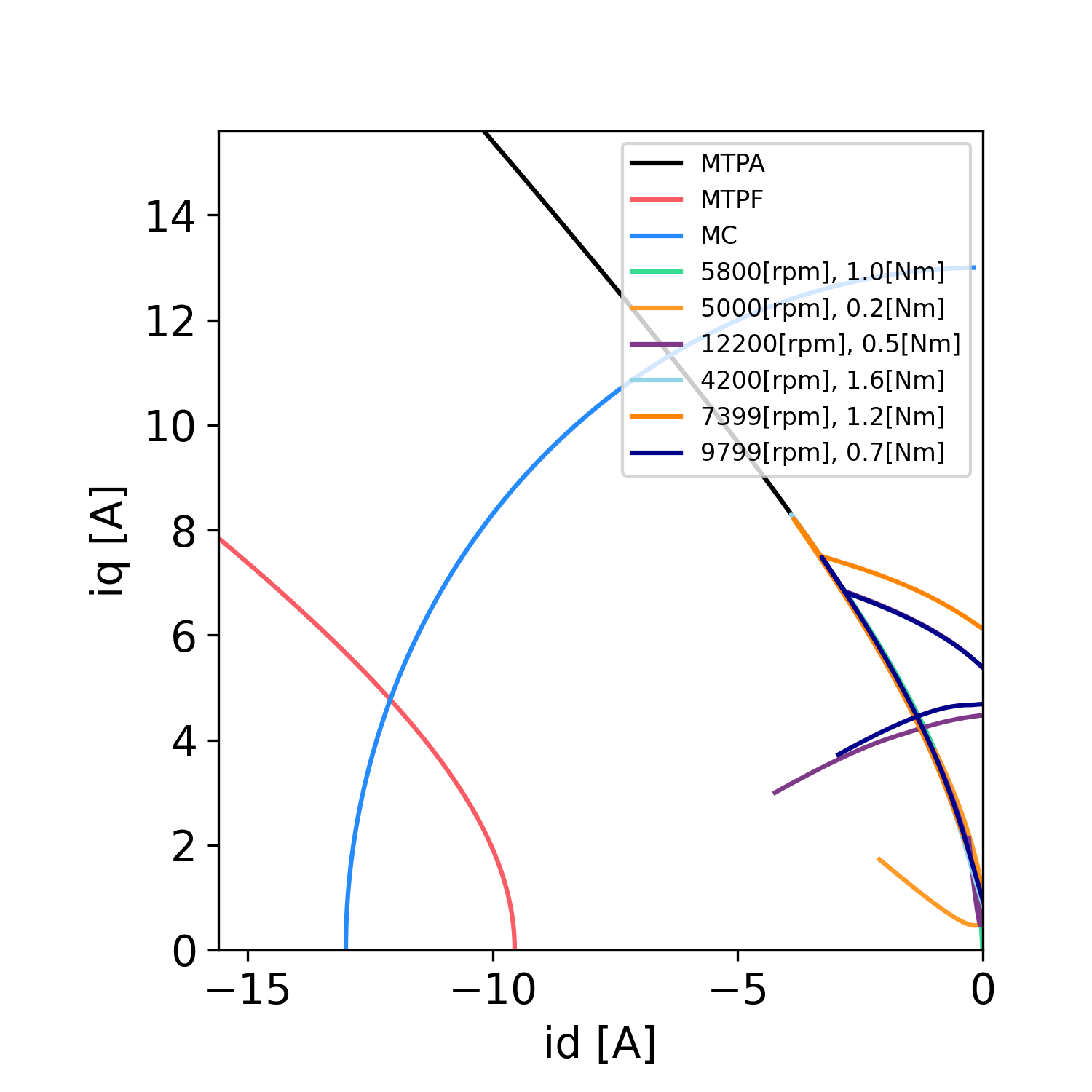}
\end{center}
\caption{Current vector space using PI-FOC (MTPA) under 1.0 s acceleration time.}
\label{fig:vec_devel_mtpa}
\end{figure}

\begin{figure}[t]
\begin{center}
  \includegraphics[width=\linewidth]{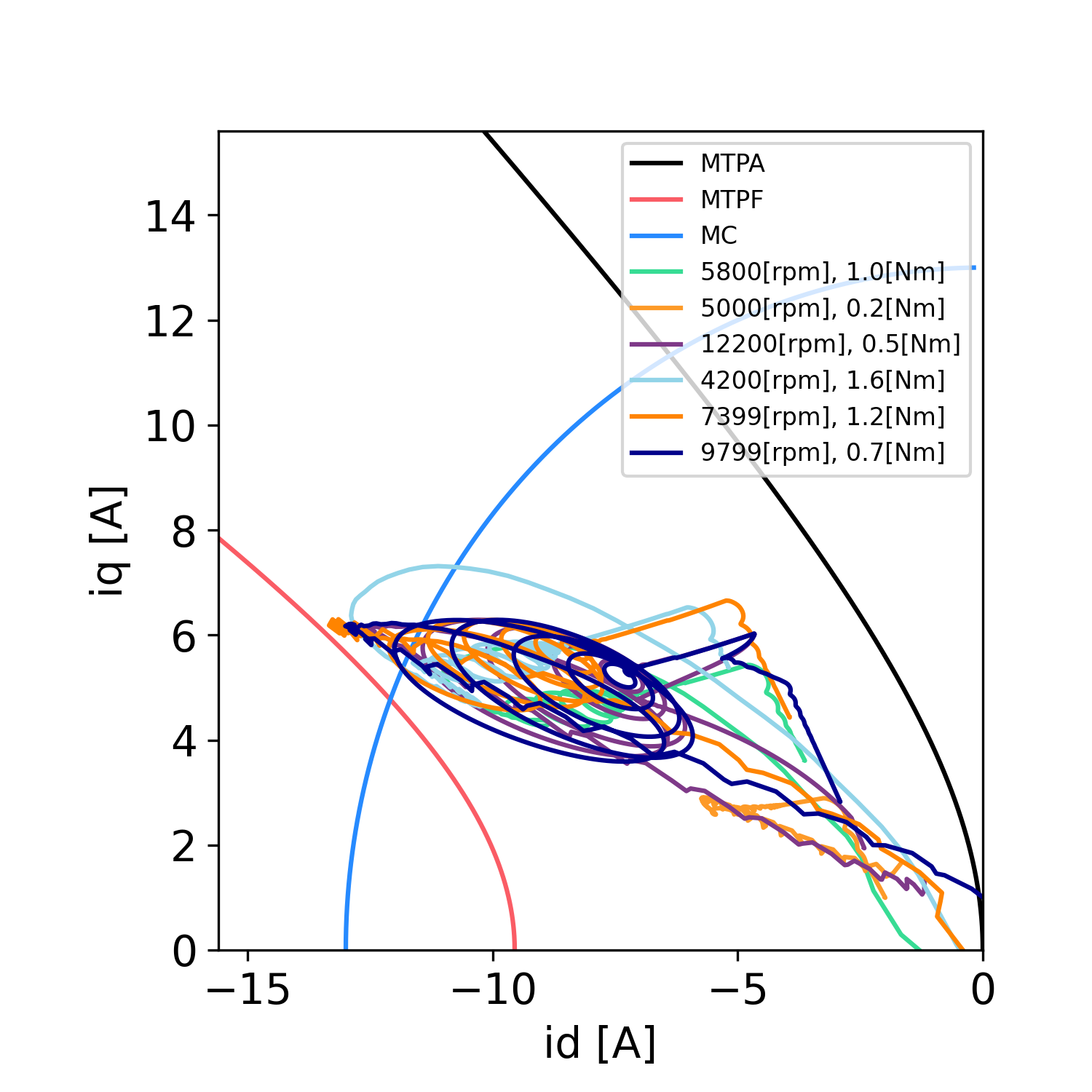}
\end{center}
\caption{Current vector space using RNN under 1.0 s acceleration time.}
\label{fig:vec_devel_rnn}
\end{figure}

To discuss the reason for the fast responses of the RNN, the current trajectories of the PI-FOC (MC) are shown in Fig. \ref{fig:vec_devel_mc}, the trajectories of the PI-FOC (MTPA) are shown in Fig. \ref{fig:vec_devel_mtpa}, and the trajectories of the RNN is shown in Fig. \ref{fig:vec_devel_rnn}. If we use PI-FOC, the current trajectory is basically limited on the selection of the reference $i_{dq}$ relationships. As the relationships are based on the steady-state assumption, the trajectory design strategy may not be optimal for transient performances. On the other hand, the current trajectories of the RNN are free from human design and actually walk through the flux-weakening region to maximize transient performances.

The remaining weakness of the RNN controller is the exact elimination of steady-state error, which can theoretically be done in any PI-FOC. However, if this property is important for user requirements, this can be easily achieved by switching the RNN and a simple PI controller, which can sustain steady-state, by the closeness of the current state and the target equilibrium.

\begin{figure}[t]
\begin{center}
 \includegraphics[width=\linewidth]{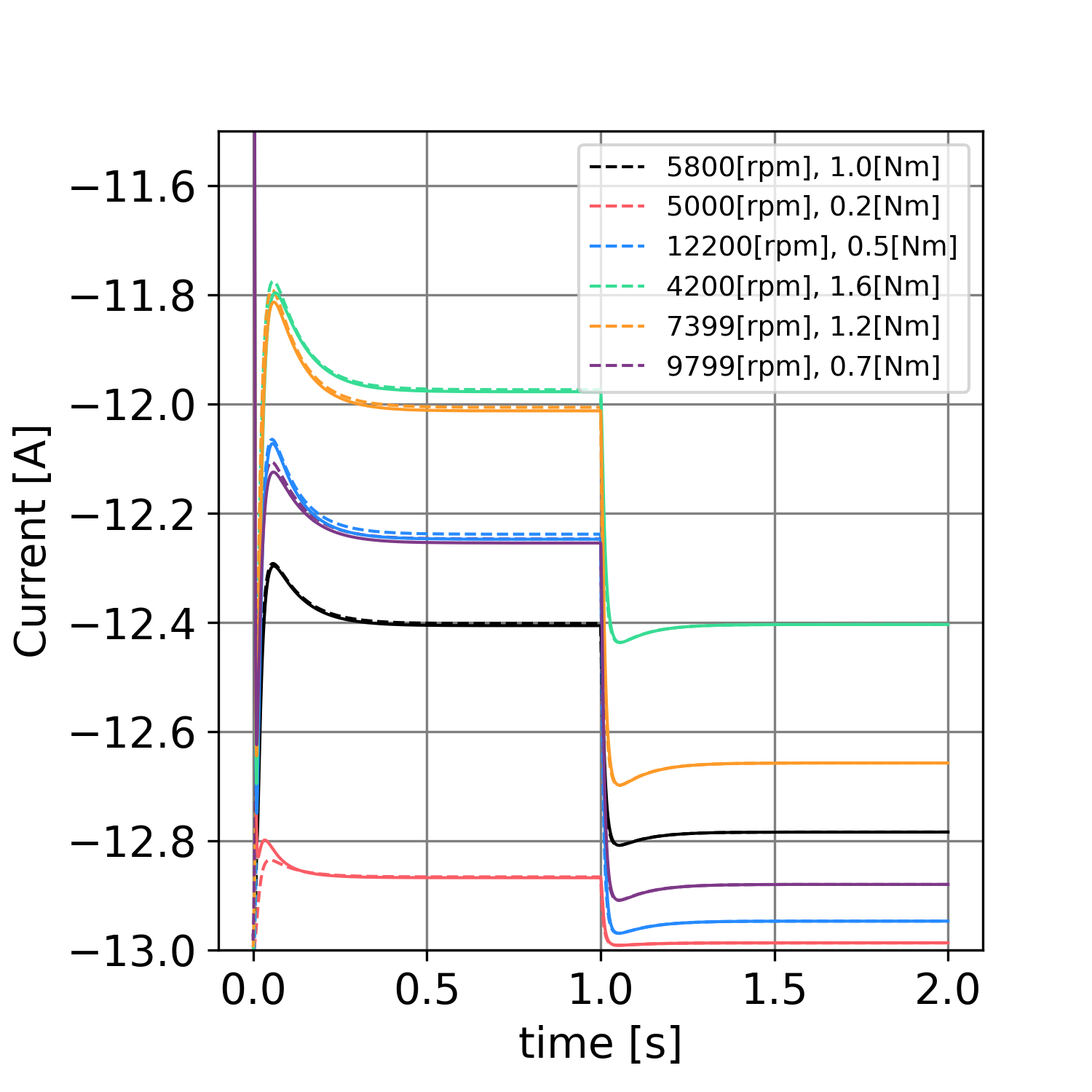}
\end{center}
\caption{d-axis current responses (solid) and references (dotted) using PI-FOC (MC) under 1.0 s acceleration time.}
\label{fig:id_devel_mc}
\end{figure}

\begin{figure}[t]
\begin{center}
 \includegraphics[width=\linewidth]{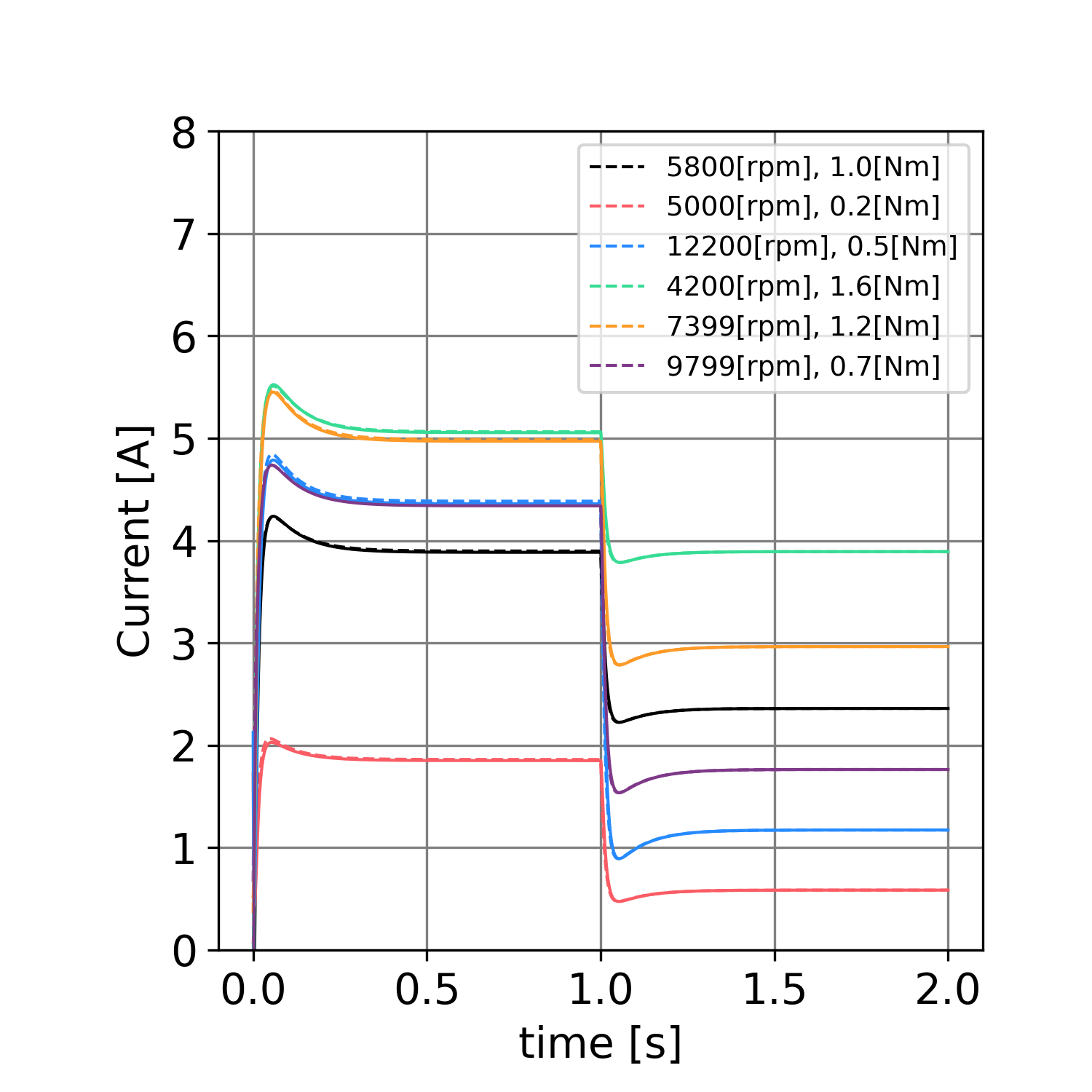}
\end{center}
\caption{q-axis current responses (solid) and references (dotted) using PI-FOC (MC) under 1.0 s acceleration time.}
\label{fig:iq_devel_mc}
\end{figure}

\begin{figure}[t]
\begin{center}
 \includegraphics[width=\linewidth]{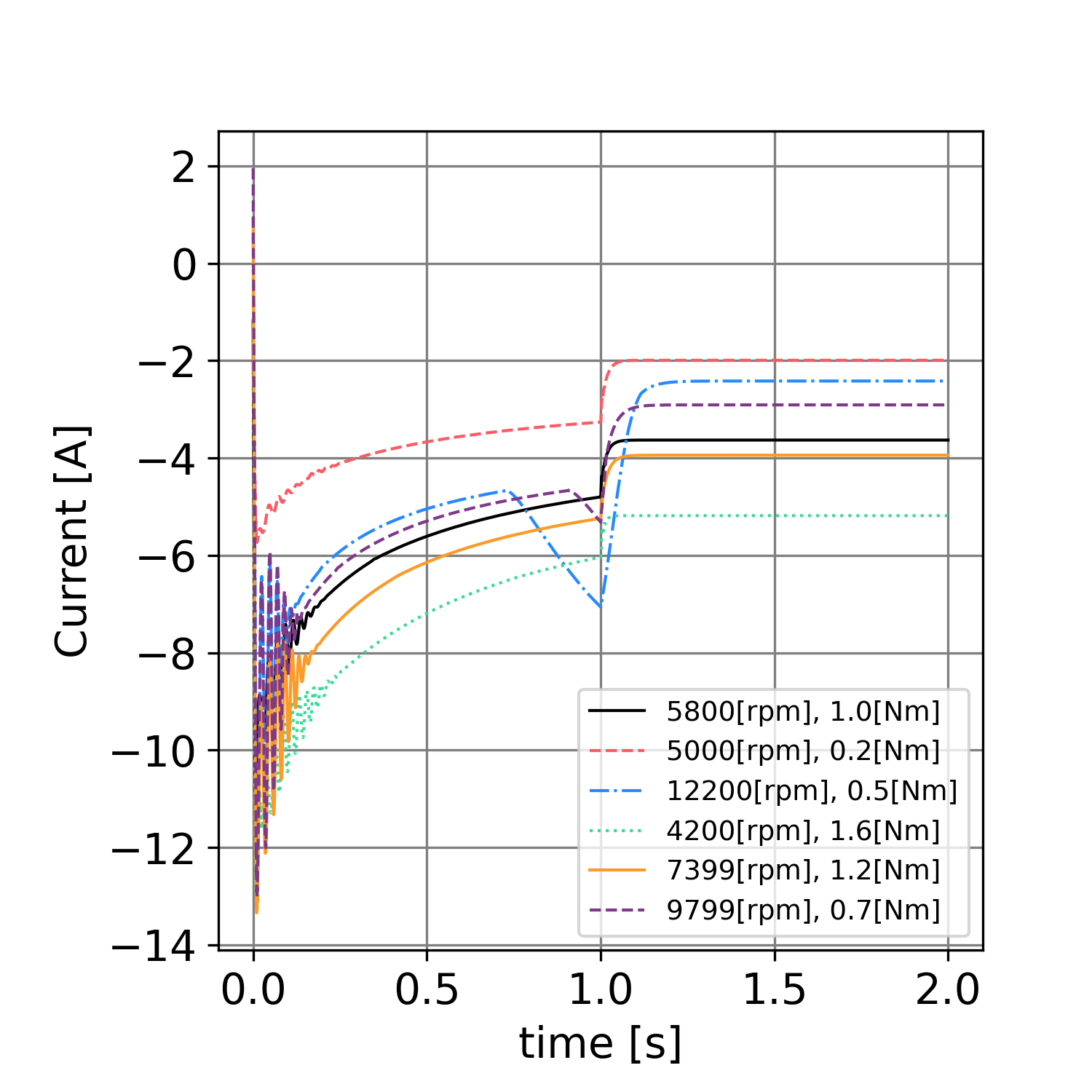}
\end{center}
\caption{d-axis current responses using RNN under 1.0 s acceleration time.}
\label{fig:id_devel_rnn}
\end{figure}

\begin{figure}[t]
\begin{center}
 \includegraphics[width=\linewidth]{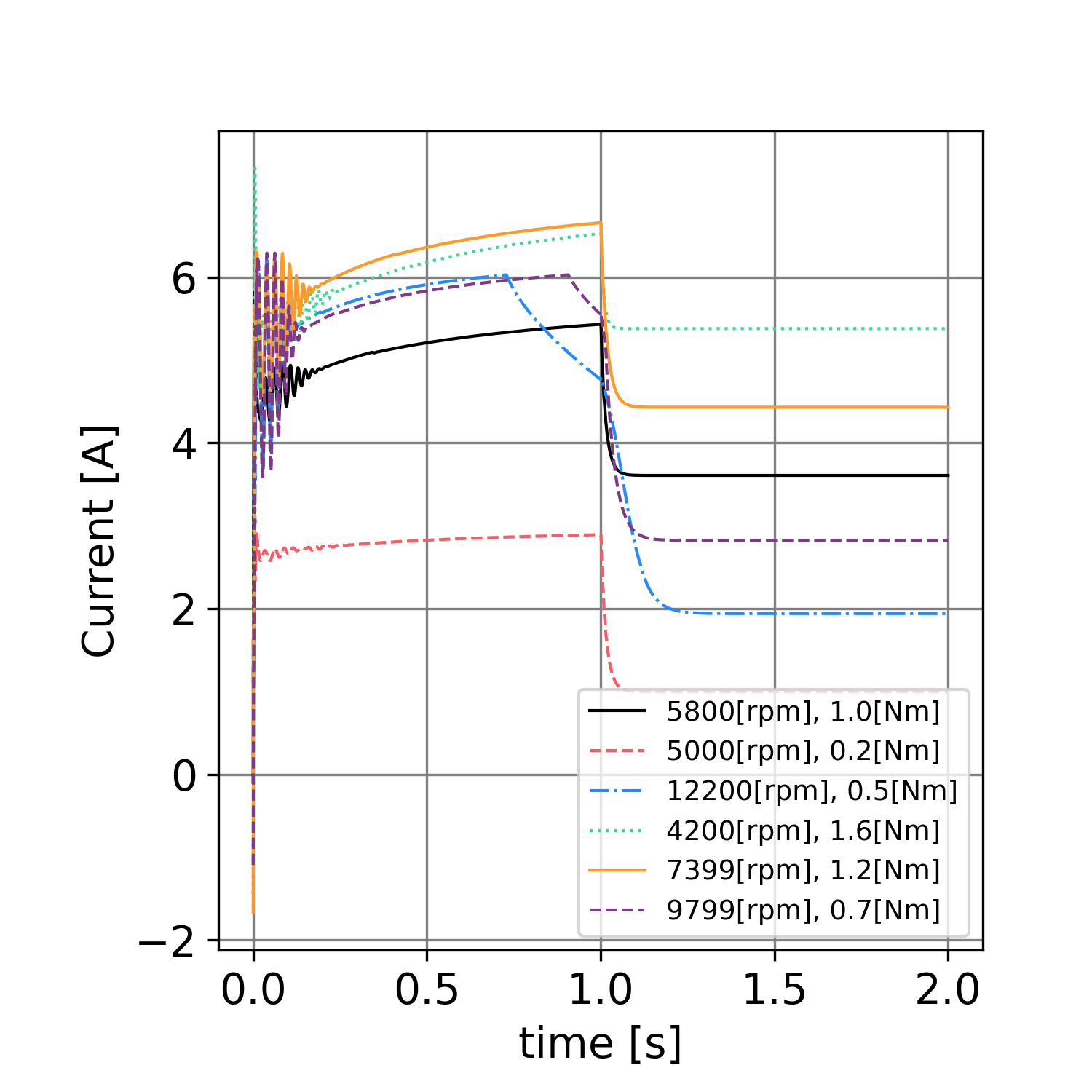}
\end{center}
\caption{q-axis current responses using RNN under 1.0 s acceleration time.}
\label{fig:iq_devel_rnn}
\end{figure}

\begin{figure}[t]
\begin{center}
 \includegraphics[width=\linewidth]{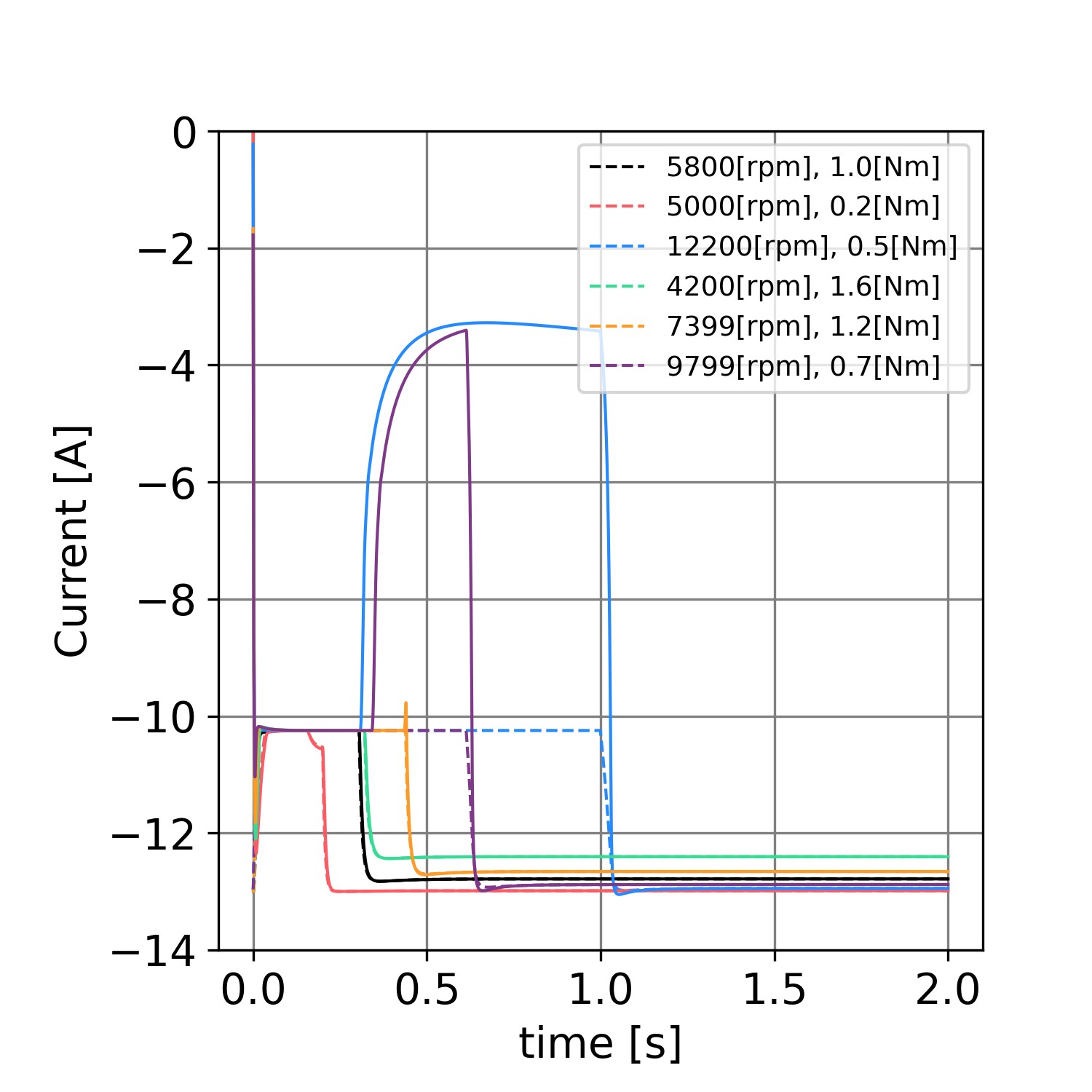}
\end{center}
\caption{d-axis current responses (solid) and references (dotted) using PI-FOC (MC with limiters) under 0.2 s acceleration time.}
\label{fig:id_ext_mc}
\end{figure}

\begin{figure}[t]
\begin{center}
 \includegraphics[width=\linewidth]{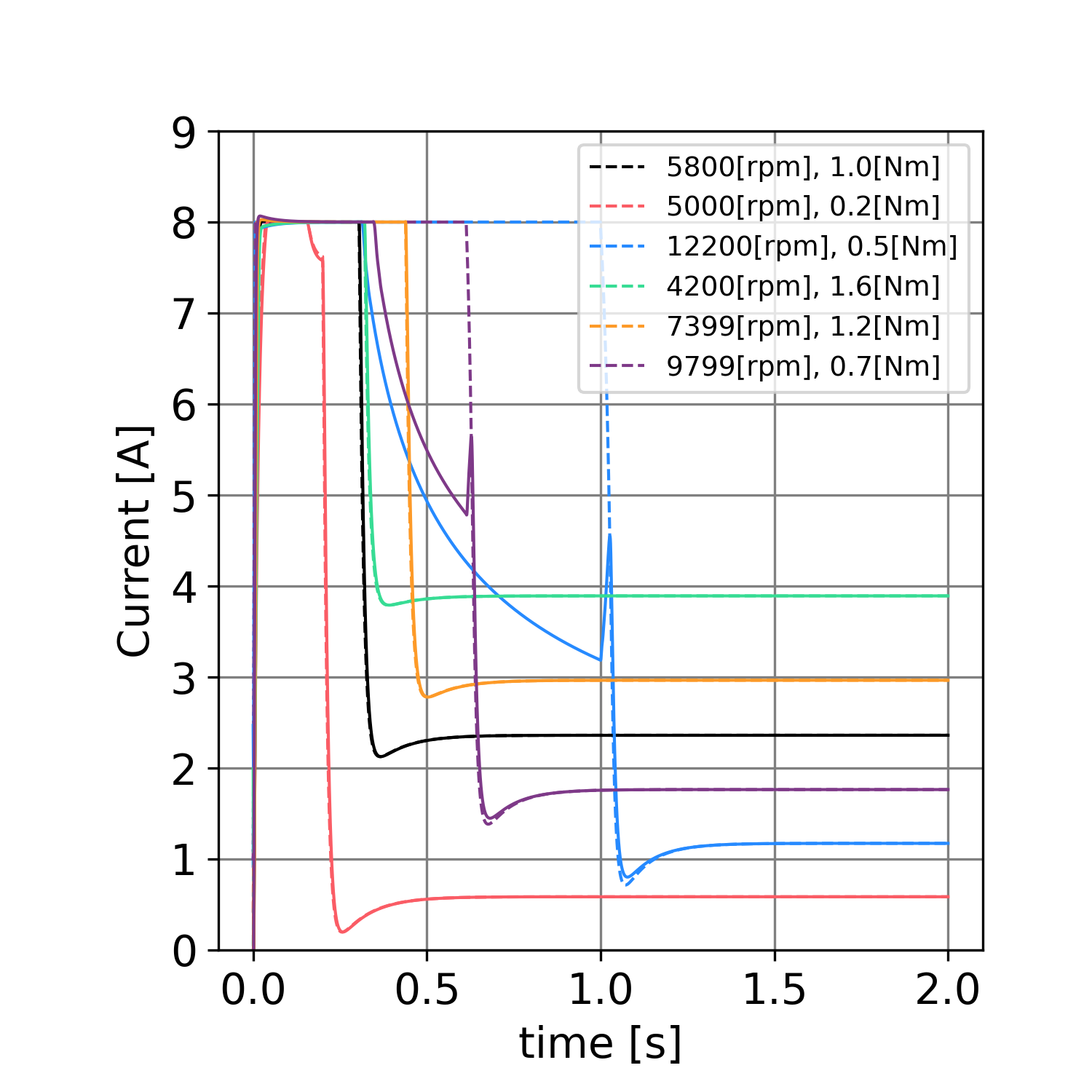}
\end{center}
\caption{q-axis current responses (solid) and references (dotted) using PI-FOC (MC with limiters) under 0.2 s acceleration time.}
\label{fig:iq_ext_mc}
\end{figure}

\begin{figure}[t]
\begin{center}
 \includegraphics[width=\linewidth]{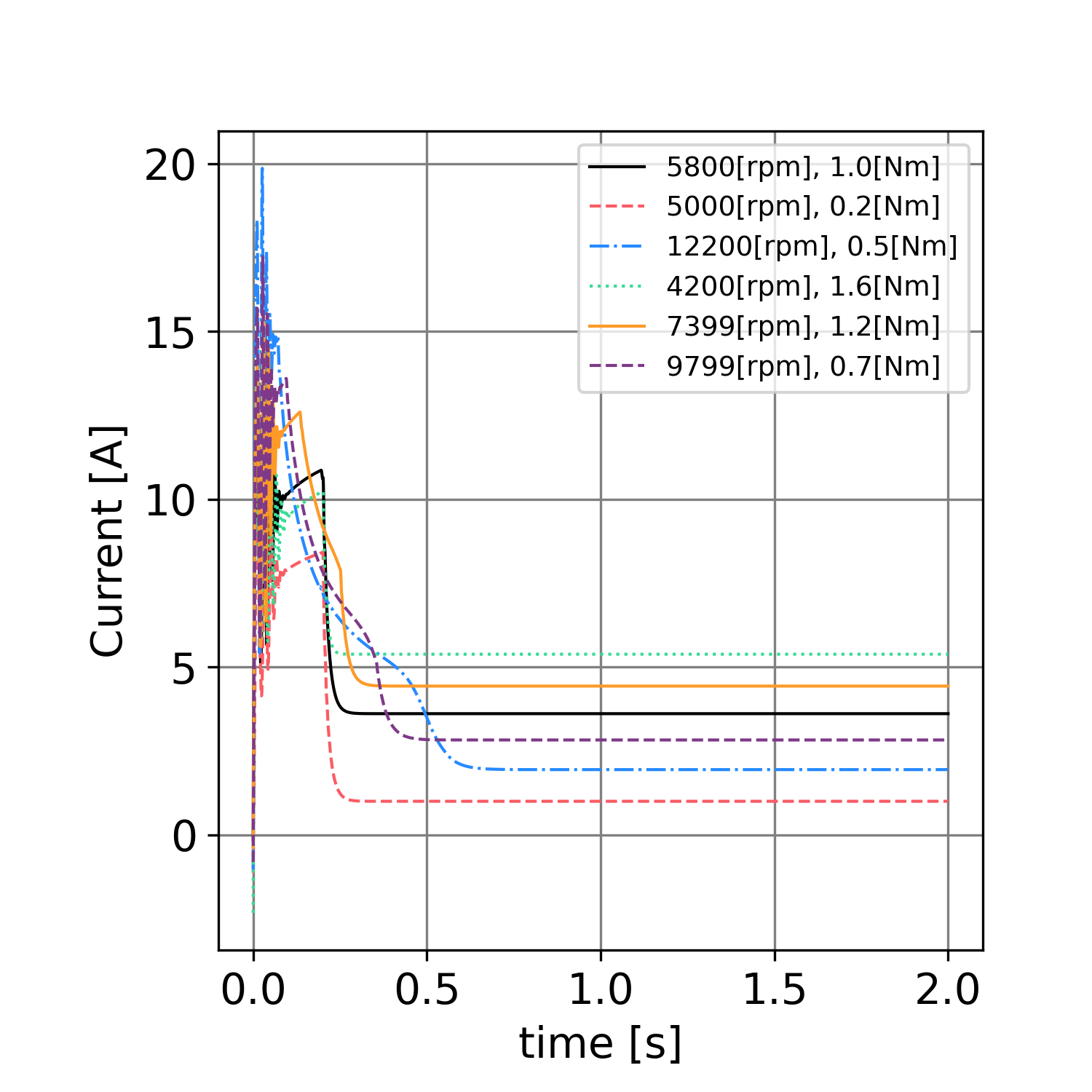}
\end{center}
\caption{q-axis current responses using RNN under 0.2 s acceleration time.}
\label{fig:iq_ext_rnn}
\end{figure}

Finally, we show the current responses of the PI-FOC(1.0 s acceleration) in Fig. \ref{fig:id_devel_mc} (d-axis), Fig. \ref{fig:iq_devel_mc} (q-axis), the RNN (1.0 s acceleration) in Fig. \ref{fig:id_devel_rnn} (d-axis), Fig. \ref{fig:iq_devel_rnn} (q-axis), the PI-FOC (0.2 s acceleration) in Fig. \ref{fig:id_ext_mc} (d-axis), Fig. \ref{fig:iq_ext_mc} (q-axis), and the RNN (0.2 s acceleration) in Fig. \ref{fig:iq_ext_rnn} (d-axis), Fig. \ref{fig:iq_ext_rnn} (q-axis). From the figures of the PI-FOC current responses, the responses resemble typical step responses. However, the RNN current responses consist of more flexible curves. As we did not set explicit current limiters in the RNN controller, some spikes can be found in the response. However, we consider these spikes do not produce real problems because transient current spikes do not produce problematic heat. Moreover, the ability to produce such current spikes contributes to the faster torque and speed response.

\begin{figure}[t]
\begin{center}
 \includegraphics[width=\linewidth]{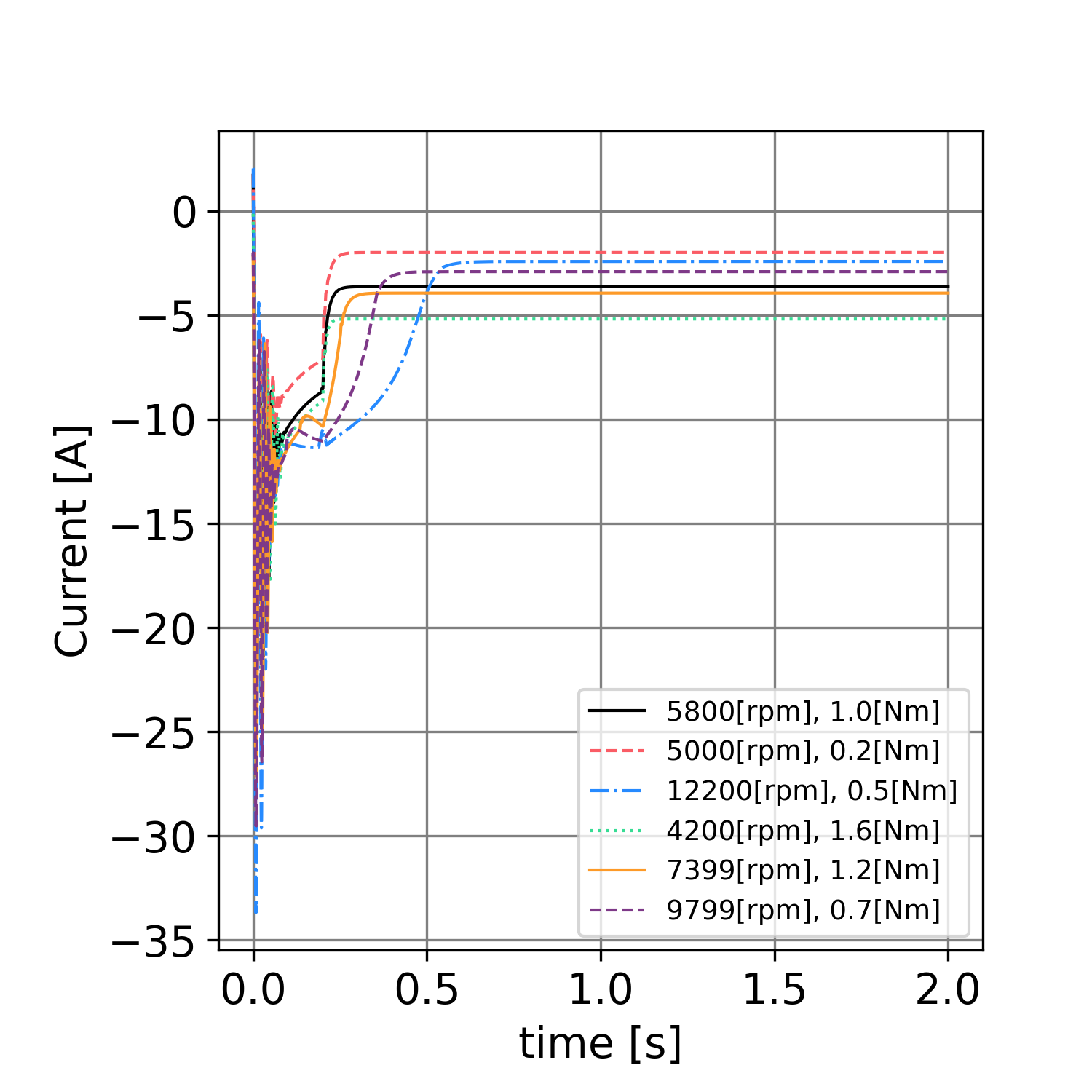}
\end{center}
\caption{d-axis current responses using RNN under 0.2 s acceleration time.}
\label{fig:id_ext_rnn}
\end{figure}

\begin{figure}[t]
\begin{center}
 \includegraphics[width=\linewidth]{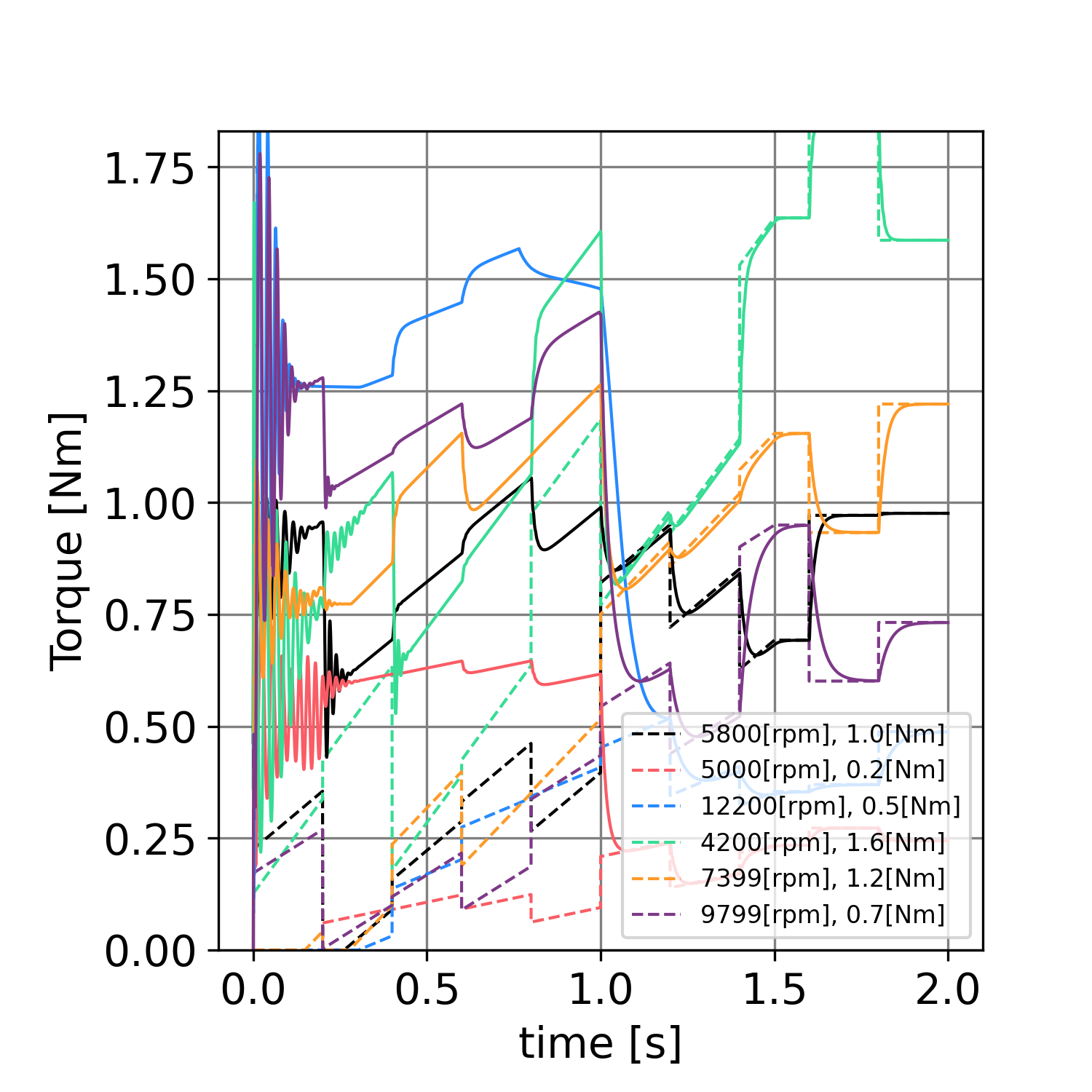}
\end{center}
\caption{Fluctuated torque functions using RNN. Dotted lines are the load torque functions. Solid lines are the electrical torque functions.}
\label{fig:TL_fluct}
\end{figure}

\begin{figure}[t]
\begin{center}
 \includegraphics[width=\linewidth]{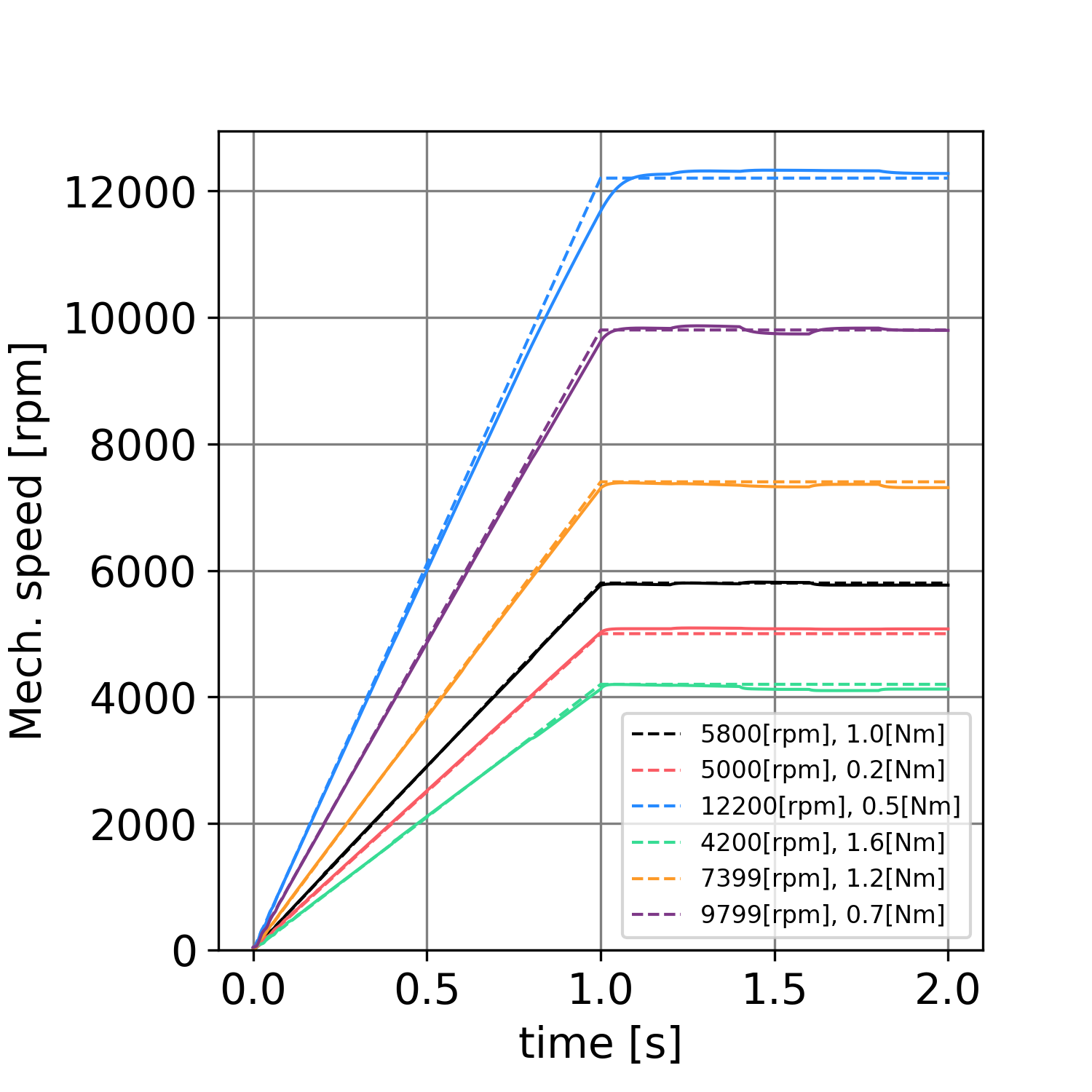}
\end{center}
\caption{Speed responses under fluctuated torque functions using RNN.}
\label{fig:TL_fluct_speed}
\end{figure}

In order to evaluate the parameter mismatch between the design and the real machine, we simulated this situation by changing the motor parameter. The result of the RNN is shown in Table \ref{table:paraint}. From the result, the RNN sustains the 2\% settling performance of approximately $\pm 20 \%$ of parameter fluctuation except in the demagnetization of the permanent magnet, which is considered a physical limitation. From an electrical perspective, we consider this performance is enough because typical electric parts have $\pm 5\%$ error.
We also evaluated the mismatch of the load torque function. We added a ramp and 30\% random fluctuation which resembles more realistic data than the step function. This setting is harder than previous work\cite{9615146}, consisting of four fixed torque constants. The torque functions are shown in Fig. \ref{fig:TL_fluct} and the speed responses are in Fig. \ref{fig:TL_fluct_speed}. From this result, the RNN controller, even trained only on step load torque function, cancels most of the load torque fluctuation.

\begin{table*}[tb]
\caption{Parameter mismatch evaluation of the RNN controller under 1.0 s acceleration time. The values are the rate of the 2\% settled simulation points in the speed-torque space that sustained from the exact parameter model within the simulation time.}
\label{table:paraint}
\begin{center}
\begin{tabular}{c|c|c|c|c|c|c|c|c|c}
\noalign{\hrule height 0.4mm}
& -50\% & -20\% & -5\% & $\pm 0 \%$ & +5\% & +20\% & +50\% & +100\% & +400\% \\
\hline
${\rm \Phi}$ &0.44 & 0.78 & 0.94& 1 & 1.00 & 1.00 & 0.99 & 0.93 & 0.00 \\
R  &0.99 & 1.00 & 1.00 & 1& 1.00 & 0.99 & 0.98 & 0.96 & 0.88 \\
$L_d$  &1.00 & 1.00 & 1.00 & 1& 0.96 & 0.89 & 0.77 & 0.11 & 0.00 \\
$L_q$  &0.54 & 0.98 & 1.00 & 1& 0.97 & 0.90 & 0.81 & 0.69 & 0.48 \\
J  &0.89 & 0.99 & 1.00 & 1& 1.00 & 1.00 & 0.99 & 0.99 & 0.81 \\
\noalign{\hrule height 0.4mm}
\end{tabular}
\end{center}
\end{table*}

\section{Conclusion}
\label{sec:conc}
We introduced an end-to-end Lipschitz-regularized RNN controller to minimize the speed transient response. We compared the RNN and a common PI-FOC with the maximum current criterion, which is supposed to have almost the widest operation range in the speed-torque space. From a computer simulation, our RNN controller, trained with an acceleration of time 1 s, and the PI-FOC were comparative in terms of the settling time, and the RNN had nearly optimal copper loss similar to the MTPA criterion. Moreover, our RNN controller extrapolated well at a quicker acceleration time of 0.2 s and maintained an acceptable copper loss, while the PI-FOC showed speed transient performance degradation at a higher speed. The main limitation and future work are below.

\subsection{Data}
Although we used simple synthetic data (e.g., speed reference or load torque) on training, the proposed RNN controller canceled complex fluctuation of load torque on evaluation. However, for practical usage, one can correct the values from the real cars or vehicles {to} improve real world performances through fine-tuning procedure.

\subsection{Iron loss}
Iron loss, which is known to be difficult to model, is ignored in the standard PMSM ODE model. In this paper, we intended to provide an evaluation of a well-known standard model, so we did not include any kind of iron loss effect. However, our approach is plant-independent, so any type of iron loss model based on ODE can be combined with the plant model. For example, there are some hysteresis models written in ODEs \cite{song2006generalized}. This attempt will provide a stricter evaluation of motor efficiency at high speed.

\subsection{System identification}
The proposed controller showed robustness on approximately $\pm20\%$ electrical or mechanical parameter variation. However, related to iron loss, the strict plant modeling itself {is} difficult. As the plant model is just a vector field of phase velocity, it can also be modeled in an ML way using its responses as training data.

\subsection{Stochasticity}
Stochasticity was not explicitly considered in this paper. Although a large number of successful deep learning techniques do not explicitly consider stochasticity and minibatch dynamics give the controller robustness to some extent, an exact formulation of stochasticity may increase robustness. A probabilistic counterpart of the ODEs is the so-called stochastic differential equations (SDEs). For example, it is known that we can derive a continuous type of evidence (the marginal likelihood) lower bound for It\^{o} SDEs\cite{huang2021variational}. This can be used to extend this paper to more precisely deal with system stochasticity.

\subsection{Inverter modeling}
The actual motors run using a three-phase (UVW) voltage. Transformation between the UVW space and the dq space using the absolute angle is required but omitted in this work. We assumed that the absolute angle is known, but an explicit treatment will be important, especially in so-called sensor-less operation, which omits the rotational sensor. Furthermore, the inverter is said to be digital because it is governed by the binary states of six switching semiconductors, which is considered to be a three-phase H-bridge circuit, and actually mimics analog control using Pulse Width Modulation (PWM) techniques. In this paper, we assumed the ideal analog control, which is operated by an analog amplifier. The digital treatment of the controller should be more difficult because digital values cannot be differentiated straightforwardly, but negative effects of the PWM, e.g., the minor loop problem\cite{Barbisio2004minor, Kawabe2012Minor}, may be modeled through this approach.

\section*{Acknowledgment}
We thank our colleague Teppei Suzuki who discussed the PyTorch implementation of the system.

\bibliography{main}
\bibliographystyle{abbrv}

\end{document}